\documentclass[pop,superscriptaddress,showpacs,amsmath,amssymb,reprint]{revtex4-1}
\usepackage{graphicx,dcolumn,bm,amsmath}
\usepackage[normalem]{ulem}

\newcommand{\beq}{\begin{equation}}
\newcommand{\enq}{\end{equation}}

\newcommand\Real{\mbox{Re}} 
\newcommand\Imag{\mbox{Im}} 

\newcommand{\e}{\eqref}

\newcommand \bfr{{\bf r}}
\newcommand \bfk{{\bf k}}

\newcommand{\p}{\partial}

\DeclareMathOperator\erfi{erfi}

\begin{document}

\title{
Langmuir wave filamentation in the kinetic regime. I. Filamentation instability of Bernstein-Greene-Kruskal modes in multidimensional Vlasov simulations
}

\author{ Denis A. Silantyev}
\affiliation{Department on Mathematics and Statistics, University of New Mexico, New Mexico 87131, USA}
\author{Pavel M. Lushnikov}
\email{plushnik@math.unm.edu}
\affiliation{Department on Mathematics and Statistics, University of New Mexico, New Mexico 87131, USA}
\author{Harvey A. Rose}
\affiliation{Theoretical Division, Los Alamos National Laboratory,
  MS-B213, Los Alamos, New Mexico, 87545}
\affiliation{New Mexico Consortium, Los Alamos, New Mexico 87544, USA}

\date{
\today
}

\begin{abstract}
A nonlinear Langmuir wave in the kinetic regime $k\lambda_D\gtrsim0.2$ may have a filamentation instability, where $k$ is the wavenumber and $\lambda_D$ is the Debye length.  The nonlinear stage of that instability develops into the filamentation of Langmuir waves which in turn leads to the saturation of the stimulated Raman scattering in laser-plasma interaction experiments. Here we study the linear stage of the filamentation instability of the particular family \cite{RoseRussellPOP2001} of Bernstein-Greene-Kruskal (BGK) modes \cite{BernsteinGreeneKruskal1957} that is a bifurcation of the linear Langmuir wave. Performing direct $2+2D$ Vlasov-Poisson simulations of collisionless plasma we find the growth rates of oblique modes of the electric field as a function of BGK's amplitude, wavenumber and the angle of the oblique mode's wavevector relative to the BGK's wavevector. Simulation results are compared to theoretical predictions.

\end{abstract}


\maketitle

\section{Introduction}

Consider a Langmuir wave \cite{PitaevskiiLifshitzPhysicalKineticsBook1981,NicholsonBook1983} (LW) wave packet with a typical wavenumber $k$.  If
$k\lambda_D\lesssim0.2$, then the hydrodynamic approximation  (the ``fluid" regime) to LW dynamics is valid, where  $\lambda_D$ is the Debye length.  In that regime a LW has a nonlinear frequency shift $\Delta \omega_{fluid}$, due to electron dynamics, proportional to the squared LW electric field amplitude $E ,$ i.e. $\Delta \omega_{fluid}\propto|E|^2$ \cite{CoffeyPhysFL1971,DewarPhysFL1972,DewarLindlPhisFL1972}. As shown in Ref. \cite{RosePOP2005}, the transition from the fluid to the "kinetic" regime occurs at $k\lambda_D\sim0.2$ when trapped electron effects cannot be ignored. The LW frequency shift due to electron trapping, $\Delta \omega_{trapped}, $ perturbatively varies as
  $\Delta \omega_{trapped}\propto|E|^{1/2}$ \cite{ManheimerFlynnPhysFL1971,DewarPhysFL1972,MoralesNeilPRL1972,RoseRussellPOP2001,RosePOP2005} with possible higher order corrections as discussed in Ref. \cite{ChangDodinPOP2015}. Thus $\Delta \omega_{trapped}$ at $k\lambda_D\gtrsim0.2$ may dominate \cite{RosePOP2005,KlineMontgomeryBezzeridesCobblePRL2005,KlineMontgomeryYin2006,WinjumFahlenMoriPOP2007,ChangDodinPOP2015} over $\Delta \omega_{fluid}$. Negative $\Delta \omega_{trapped}$, with positive diffraction, imply  LW filamentation \cite{RosePOP2005,RoseYinPOP2008,YinAlbrightBowersDaughtonRosePOP2008,BergerBrunnerBanksCohenWinjumPOP2015}.      $3D$  particle-in-cell (PIC) \cite{BowersAlbrightYinBergenKwanPOP2008} simulation results have been interpreted as
showing that the trapped electron LW filamentation instability can saturate \cite{YinAlbrightRoseBowersPOP2009,YinAlbrightRoseMontgomeryKlineKirkwoodMilovichFinneganBergenBowersPOP2014}
stimulated Raman back-scatter (SRS)  \cite{GoldmanBoisPhisFL1965} by reducing the LW's coherence. In actual
plasma, the SRS daughter LW is subject to other instabilities as well, such as LW-ion-acoustic
decay (LDI). Fluid and kinetic regime LDI have been observed in SRS
simulations \cite{YinAlbrightRoseBowersPOP2009,RussellDuBoisRosePOP1999} while kinetic regime LDI has been experimentally \cite{MontgomeryCobbleFernandezFociaPOP2002} noted \cite{KlineMontgomeryBezzeridesCobblePRL2005}.

Additional complexity in the interpretation of experimental data gathered from laser-plasma
interaction arises from instabilities of the laser beam \cite{MaxAronsLangdonPRL1974,DepierreuxNature2014} coupled \cite{SchmittAfeyanPOP1998,LushnikovRosePlasmPysContrFusion2006,LushnikovRosePRL2004} to
relatively low frequency ion-acoustic waves. Since direct experimental data pointing
to kinetic LW filamentation has not been available, first principles simulation of pure
LW dynamics is perhaps the cleanest way to ``see" this phenomenon. Fully nonlinear
PIC simulations \cite{YinAlbrightBowersDaughtonRosePOP2008} with Bernstein-Greene-Kruskal (BGK) mode \cite{BernsteinGreeneKruskal1957} initial conditions have shown qualitative
agreement with LW filamentation theory \cite{RosePOP2005,RoseYinPOP2008}, but the theory''s finer points, such as
instability thresholds, require a noise free model, namely the Vlasov simulations.

Here we address LW filamentation in the kinetic regime with $k\lambda_D>0.3$ by studing the filamentation instability of BGK modes using $2+2D$ (two velocity and two spatial dimensions) spectral Vlasov simulations. Our simulations only include collisionless electrostatic electron dynamics in a static neutralizing ion background, thereby excluding  the LW ion-acoustic decay and ponderomotive LW filamentation instabilities, amongst others. BGK modes are constructed following  the approach of Ref. \cite{RoseRussellPOP2001} to approximate the adiabatically slow pumping by SRS. We concentrate on the linear stage of the filamentation instability development while observing strong LW filamentation in the nonlinear stage. Also in the second paper (Part II) of the series,  we consider dynamically prepared BGK-like initial conditions created with slow SRS-like pumping   (similar to Ref. \cite{BergerBrunnerBanksCohenWinjumPOP2015})  and study the filamentation instability of those waves comparing both with  the results of this paper for BGK modes and the results of Ref. \cite{BergerBrunnerBanksCohenWinjumPOP2015}.

The paper is organized as follows. Section \ref{sec:Basicequations} introduces the Vlasov-Poisson system and its general BGK solutions (equilibria). In Section \ref{sec:1DEquilibrium}
we recall a special family \cite{RoseRussellPOP2001} of $1+1D$ BGK modes that bifurcate from linear LW. We describe the analytical and numerical construction of these modes. Section \ref{sec:DispersionRelation} outlines their nonlinear dispersion relation and Section \ref{sec:TransverseInstability} provides filamentation's definition and analytical results on its growth rate. In Section \ref{sec:NUMERICALSIMULATIONS} we provide results of $2+2D$ Vlasov simulations.  Section \ref{sec:Simulationsettings} is devoted to the Vlasov simulations settings and our numerical method. Section \ref{sec:2DHarveyBGK} addresses filamentation instability results and their comparison with theory.
 Section \ref{sec:PICComparison} provides a comparison of the growth rates obtained in Section \ref{sec:2DHarveyBGK}  with the growth rates  from PIC code simulations of Ref. \cite{YinAlbrightBowersDaughtonRosePOP2008}.
 In Section \ref{sec:Conclusion} the main results of the paper are discussed.

\section{Basic equations}
\label{sec:Basicequations}

The Vlasov equation for the phase
space distribution function $f(\bfr,{\bf v},t)$, in units such that electron mass $m_e$ and charge $e$ are
normalized to unity, the spatial coordinate $\bfr=(x,y,z)$ to the electron Debye length $\lambda_D$, the time $t$ to  reciprocal electron plasma
frequency, $1/\omega_{pe}$, \cite{RoseDaughtonPOP2011} and  the velocity ${\bf v}=(v_x,v_y,v_z)$ is normalized to the the electron
thermal speed $v_e$, is
\begin{equation}
    \left\lbrace\frac{\partial }{\partial t} + {\bf v}\cdot\nabla+ {\bf E}\cdot \frac{\partial }{\partial {\bf v}} \right\rbrace f=0,
    \label{eq:vlasov}
\end{equation}
where $\bf E$ is the electric field scaled to $k_BT_e/(\lambda _De).$   Here  $T_e$ is the background electron temperature and $k_B$  is the Boltzmann constant. Magnetic field effects are ignored for clarity. Then, in the electrostatic regime,
\begin{align} \label{phidef}
{\bf E}=-\nabla \Phi,
\end{align}
 with  the electrostatic potential $\Phi$ given by Poisson's equation
\begin{align} \label{Poisson}
\nabla^2\Phi=1-\rho,
\end{align}
and electron density, $\rho$, is given by
\begin{equation}
    \rho ({\bf r},t)=\int  f({\bf r},{\bf v},t)  d{\bf v}.
    \label{eq:density}
\end{equation}
The usual factor of $4\pi$ is absent from equation \eqref{Poisson} because of the chosen normalization and 1 in equation \eqref{Poisson} comes from the neutralizing ion background.

Equations \e{eq:vlasov}-\e{eq:density} form the closed Vlasov-Poisson system. Its finite amplitude  travelling wave solutions, moving with phase velocity, $v_\varphi$, are called Bernstein-Greene-Kruskal (BGK) modes \cite{BernsteinGreeneKruskal1957}. Here we assume without loss of generality that $z$ is chosen in the direction of $v_{\varphi}$ so that $f$ assumes the form $f(\bfr_\perp,z-v_{\varphi}t,{\bf v})$, with $\bfr_\perp\equiv(x,y)$, and equation \e{eq:vlasov} reduces to
\begin{equation}
    (v_z-v_{\varphi})\frac{\partial }{\partial z} f+ {\bf v}_\perp\cdot\nabla f+ {\bf E}\cdot \frac{\partial }{\partial {\bf v}}  f=0.
    \label{eq:vlasovBGK}
\end{equation}

The general solution of equations \e{phidef} and \e{eq:vlasovBGK} is given by $f=g\left (W\right )$, where $g$ is an arbitrary function of the single scalar argument
\begin{equation} \label{Wdef}
W\equiv\frac{(v_z-v_\varphi)^2}{2}+  \frac{{\bf v}_\perp^2}{2}+\Phi(\bfr_\perp,z-v_\varphi t)
\end{equation}
  which is the single particle energy  (kinetic energy in the moving reference frame plus electrostatic energy).

BGK modes are obtained if we require $g(W)$ to satisfy  equations \e{Poisson} and \e{eq:density} \cite{BernsteinGreeneKruskal1957}.
That requirement still allows a wide variety of solutions.

\section{BGK MODE LINEAR FILAMENTATION INSTABILITY}
\label{sec:LINEARFILAMENTATIONINSTABILITY}
Our goal is to study the transverse stability of BGK modes. In general, a linear instability is specific to a given  BGK mode. We choose a BGK mode that is dynamically selected (at least approximately) by SRS with $z$ being the direction of laser beam propagation in plasma. The simplest BGK family has a nontrivial solution $f_{BGK}$  in $1+1D$ (one space and one velocity dimension \cite{BernsteinGreeneKruskal1957}) with no dependence on the transverse coordinate $\bfr_\perp,$ while the dependence on the transverse  velocity ${\bf v}_\perp$ being trivially Maxwellian as follows
\begin{equation}
    f =f_{BGK}(z-v_{\varphi}t,{v_z})\frac{\exp(-{\bf v}_\perp^2/2)}{{2\pi}}.
    \label{eq:BGKMaxwellian}
\end{equation}

Our initial model \cite{RoseRussellPOP2001} of the SRS daughter LW in a laser speckle is presented in Eq. (\ref{eq:1Dmodel}) below. If a time-dependent Vlasov equation solution has a symmetry, e.g., in $2+2D$ (two space and two velocity dimensions) when the initial condition (and possible external potential) only depend on one spatial coordinate $z$, or in $3+3D$ a cylindrically symmetric configuration, then an instability may break that symmetry, allowing for a determination of growth rate. The former, revisited here, was explored in $2+2D$ Vlasov simulations \cite{BergerBrunnerBanksCohenWinjumPOP2015}, while the latter was observed \cite{YinAlbrightRoseBowersPOP2009} in $3D$ PIC, SRS single speckle simulations. In addition, we present LW filamentation growth rates of linear fluctuations about a particular class of BGK modes, recalled in the next Section \ref{sec:1DEquilibrium}.

\subsection{Construction of 1+1D BGK }
\label{sec:1DEquilibrium}

The beating of laser and SRS light provides a source of LWs thus pumping BGK modes. Following Ref. \cite{RoseRussellPOP2001}, we assume that the laser intensity is just above SRS instability threshold. Then the pumping of LWs is slow and can be idealized as a travelling
wave sinusoidal external potential $\Phi_{ext}$, with amplitude $\phi_{pump}$, phase speed $v_\varphi$ and
 wavenumber $k_z$ such that
\begin{equation}
    \Phi_{ext} =\phi_{pump} \cos[k_z (z-v_\varphi t)], \ k_z = |{\bf k}|.
    \label{eq:EXTPotential}
\end{equation}

The total electrostatic potential, $\Phi$, is given by
\begin{equation}
    \Phi =\Phi_{ext} + \Phi_{int},
    \label{eq:ElectricPotential}
\end{equation}where the internal potential $\Phi_{int}$ is determined from Poisson's equation \e{eq:Poisson1D}, where $f_{1D}(z,v_z,t)$ is the $1D$ electron phase
space distribution function.
\begin{equation}
    \frac{\partial^2\Phi_{int}}{\partial z^2} =1-\int f_{1D}dv_z,
    \label{eq:Poisson1D}
\end{equation}

Inertial confinement fusion applications require a dynamic laser beam smoothing \cite{KatoMimaMiyanagaPRL1984,LindlPhysPlasm2004,MeezanEtAlPhysPlasm2010} resulting in a time-dependent speckle field of laser intensity. $\Phi_{ext}$ attains a local maximum in a laser speckle, which is a local maximum of laser beam intensity. Intense speckles have a width approximately $F\lambda_0$,
with $F$ the optic $f$-number (the ratio of
the focal length of the lens divided by the lens diameter) and $\lambda_0$ the laser
wavelength. The temporal scale $t_c$ of beam smoothing  is typically large compared with the inverse growth rate $1/\gamma_{SRS}$ of SRS (e.g. for the National Ignition Facility \cite{LindlPhysPlasm2004,MeezanEtAlPhysPlasm2010} $t_c\sim 4$ps and typically $1/\gamma_{SRS}\sim0.03$ps).
It implies that the speckle can be considered as time-independent which we assume below.   Electrons, with the typical speed $v_e$,  cross a
speckle's width in a dimensional time scale $1/\nu_{SideLoss} \propto F\lambda_0/v_e$.  As a result, $f_{1D}$ tends to relax to the
background distribution function, $f_0$, assumed Maxwellian,
\begin{equation}
    f_0(v_z) =\frac{\exp(-v_z^2/2)}{\sqrt{2\pi}}
    \label{eq:Maxwellian}
\end{equation}
at the rate $\nu_{SideLoss}$. These considerations motivate our $1+1D$ model of BGK generation by introducing the relaxation term $-\nu_{SideLoss}[f_{1D}(z,v_z,t)-f_0(v_z)$] into the Vlasov equation \e{eq:vlasov} as follows. In the wave frame (switching to that frame implies $z \to z +v_\varphi t$ and $v_z\to v_z+v_\varphi$),
\begin{equation}
\begin{split}
    &\left\lbrace\frac{\partial }{\partial t} + v_z\frac{\partial }{\partial z}- \frac{\partial \Phi}{\partial z} \frac{\partial }{\partial v_z} \right\rbrace f_{1D}(z,v_z,t)= \\ &-\nu_{SideLoss}[f_{1D}(z,v_z,t)-f_0(v_z+v_\varphi)].
    \label{eq:1Dmodel}
\end{split}
\end{equation}
Let $f_{eq}$ be a time independent solution of Eq. (\ref{eq:1Dmodel}). In the double limit
\begin{equation}
    f_{BGK}=\lim_{\phi_{pump}\to 0}\lim_{\nu_{SideLoss}\to 0} f_{eq},
    \label{eq:DoubleLim}
\end{equation}
a particular BGK mode which bifurcates \cite{HollowayDorningPhysRevA1991,BuchananDorningPhysRevE1995} from a linear LW, $f_{BGK}$, may be obtained \cite{RoseRussellPOP2001}. This mode correspond to the adiabatically slow pumping by SRS. It depends on
$(z,v_z)$ only through the single particle energy, $W$,
\begin{equation}
    W = \Phi(z)+v_z^2/2
    \label{eq:W}
\end{equation}
which is the restriction of equation \e{Wdef} to 1+1D case in the wave frame with $\Phi(\bfr_\perp,z-v_\varphi t)\to \Phi(z)$.

There are two methods to construct BGK modes in question. First method is numerical one and implies that we numerically solve equations  (\ref{eq:EXTPotential})-(\ref{eq:1Dmodel}) for each values of $\phi_{pump}$ and $\nu_{SideLoss}$ followed by taking numerically the double limit \e{eq:DoubleLim}.
Second method is analytical one and is based on the integration along the particle orbits of the time independent solution of Eq. (\ref{eq:1Dmodel}), where the double limit \e{eq:DoubleLim} is evaluated analytically.
We investigated both methods, found that they give similar results, but choose below to focus on the second method only since it is a simpler to implement and free of numerical issues.

The electrostatic potential $\Phi$ traps electrons with velocities close enough to $v_\varphi$ such that they cannot go over barriers created by $\Phi.$  Thus for different electrons there are both passing orbits outside the trapping region and periodic orbits inside the trapping region. Recall that passing orbits can have either positive or negative velocities, and this must be specified along
with $W$.
It was shown in Refs.  \cite{ONeilPhysFL1965} and \cite{RoseRussellPOP2001} that taking the double limit \e{eq:DoubleLim} in the equation Eq. (\ref{eq:1Dmodel}) we get
\begin{equation}
    f_{BGK}(W) = \oint_{W}f_0[v(s)+v_\varphi]ds/T(W).
    \label{eq:fBGK}
\end{equation}
The integral sign here denotes integration around a particular orbit with constant $W$. The time-like characteristic variable $s$, used in integration, parametrizes  a particular orbital location $(z(s),v(s))$ through the characteristic equations
\begin{equation}
    dz/ds = v, \ dv/ds = -d\Phi/dz.
    \label{eq:Orbit}
\end{equation}
Also $T(W)$ denotes the orbit's period,
\begin{equation}
    T(W) = \oint_{W} ds.
    \label{eq:OrbitPeriod}
\end{equation}
Here and throughout the remaining part of Section \ref{sec:LINEARFILAMENTATIONINSTABILITY} we replace $v_z$ by $v(.)$ when it describes the velocity of a particular electron with energy $W$ as a function of some parameter ($s$ or $z$), while we think of $v_z$ as independent variable in the rest of the formulas. Also we abuse notation and use the same symbols  for $v$ and $f_{BGK}$ irrespective of their parametrization by different variables.
$\Phi$ is assumed periodic so that all orbits are closed by periodicity (including the passing orbits).

\begin{figure}
\includegraphics[width=3in]{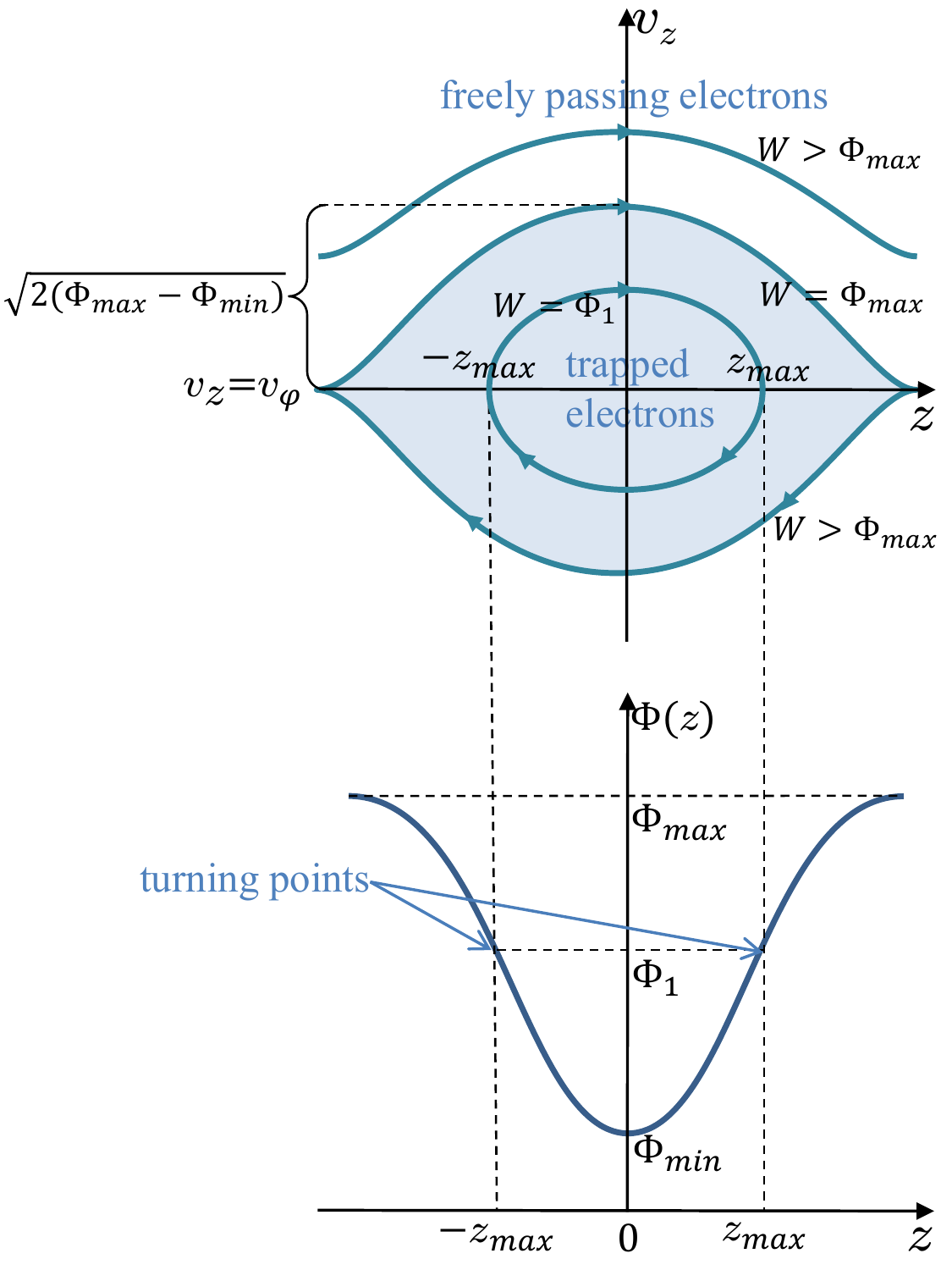}
\caption{(Color online) Schematics of the electric potential and the corresponding trapping region of $f_{BGK}(z,v_z)$.}
\label{fig:Trapping_region}
\end{figure}

Assume  $\Phi(z)$ is the given function of $z$. Then using Eq. \e{eq:W}, changing the integration variable from $s$ to $z$ in Eqs. \e{eq:fBGK} and \e{eq:OrbitPeriod},  we can express $T(W)$ and   $f_{BGK}(z,v_z)\equiv f_{BGK}(W)$ at any point $(z,v_z)$ in the phase space as follows (see Fig. \ref{fig:Trapping_region})
\begin{eqnarray} \label{Twint}
 T(W) = \begin{cases} 4\int\limits_{0}^{z_{max}} \frac{dz}{v(z)}, \ \Phi_{min}<W<\Phi_{max}, \\
    \int\limits_{0}^{L_z} \frac{dz}{v(z)}, \ W>\Phi_{max}, \end{cases} \\
\nonumber f_{BGK}(W)  T(W) = \\  \label{eq:BGK}
   \begin{cases} 2\int\limits_{0}^{z_{max}} \frac{f_0[v_\varphi+v(z)]+f_0[v_\varphi-v(z)]dz}{v(z)}, \ \Phi_{min}<W<\Phi_{max}, \\
    \int\limits_{0}^{L_z} \frac{f_0[v_\varphi+v(z)]dz}{v(z)},  \ W>\Phi_{max}\mbox{ and } v_z>v_\varphi, \\
    \int\limits_{0}^{L_z} \frac{f_0[v_\varphi-v(z)]dz}{v(z)},  \ W>\Phi_{max}\mbox{ and } v_z<v_\varphi,
    \end{cases}
\end{eqnarray}


and $v(z)$ is determined from Eq. \e{eq:W} as
\begin{equation} \label{vdef}
v(z)=\sqrt{2[W-\Phi(z)]},
\end{equation}
with $\Phi_{min}\equiv \min_z{\Phi(z)}, \ \Phi_{max}\equiv\max_z{\Phi(z), \ L_z\equiv\frac{2\pi}{k_z}}$.
We assume that $\Phi(z)$ has a single local maximum and a single local minimum per period $L_z$. Also $z_{max}$  is obtained by numerically inverting $\Phi(z_{max})=W$ (see Fig. \ref{fig:Trapping_region} for the illustration).
Note that it is also possible to integrate  Eqs. \e{eq:fBGK} and \e{eq:OrbitPeriod}
directly in $s$ variable
which has more compact form compared to Eqs. (\ref{Twint})-(\ref{eq:BGK}). However, we chose to use Eqs. (\ref{Twint})-(\ref{vdef})
because it is easier to implement a high-order numerical scheme for integrals \e{Twint} and \e{eq:BGK} that depend only on one independent variable rather than calculating integrals in Eqs. (\ref{eq:fBGK})-(\ref{eq:OrbitPeriod}) that require two-step process, first numerically finding orbits $(z(s),v(s))$ and then computing the integrals.

The amplitudes of Fourier harmonics of $\Phi(z)$  are rapidly decaying  \cite{RoseRussellPOP2001}, so we start by constructing a BGK mode approximately by taking into account only the first harmonic
\begin{equation} \label{eq:Phi}
    \Phi(z) = -\phi_0\cos(k_zz)
\end{equation}
parametrized by the amplitude $\phi_0.$
Then the comparison with definitions in Eq. \e{vdef}   implies that $ \Phi_{min}=-\phi_0, \ \Phi_{max}=\phi_0, \ v(z)=\sqrt{2(W+\phi_0\cos(k_zz))} $ and $z_{max}=\frac{1}{k_z}\arccos(\frac{-W}{\phi_0}).$

\begin{figure}
\includegraphics[width=3in]{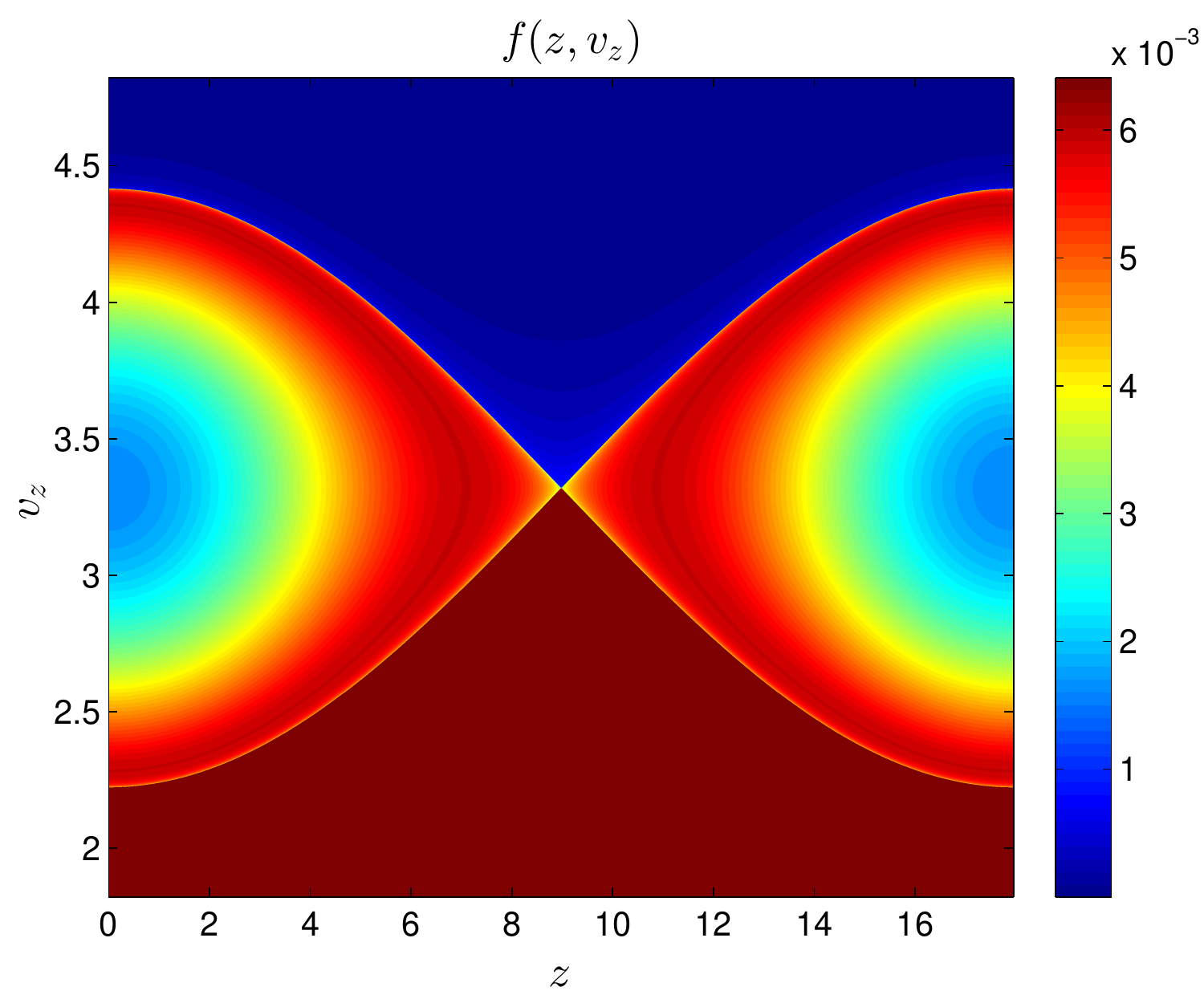}
\caption{(Color online) The phase space density distribution function $f_{BGK}(z,v_z)$ of BGK mode with $k_z=0.35, \phi_0=0.3, v_\varphi=3.321836\ldots$.}
\label{fig:BGK}
\end{figure}
\begin{figure}
\includegraphics[width=3in]{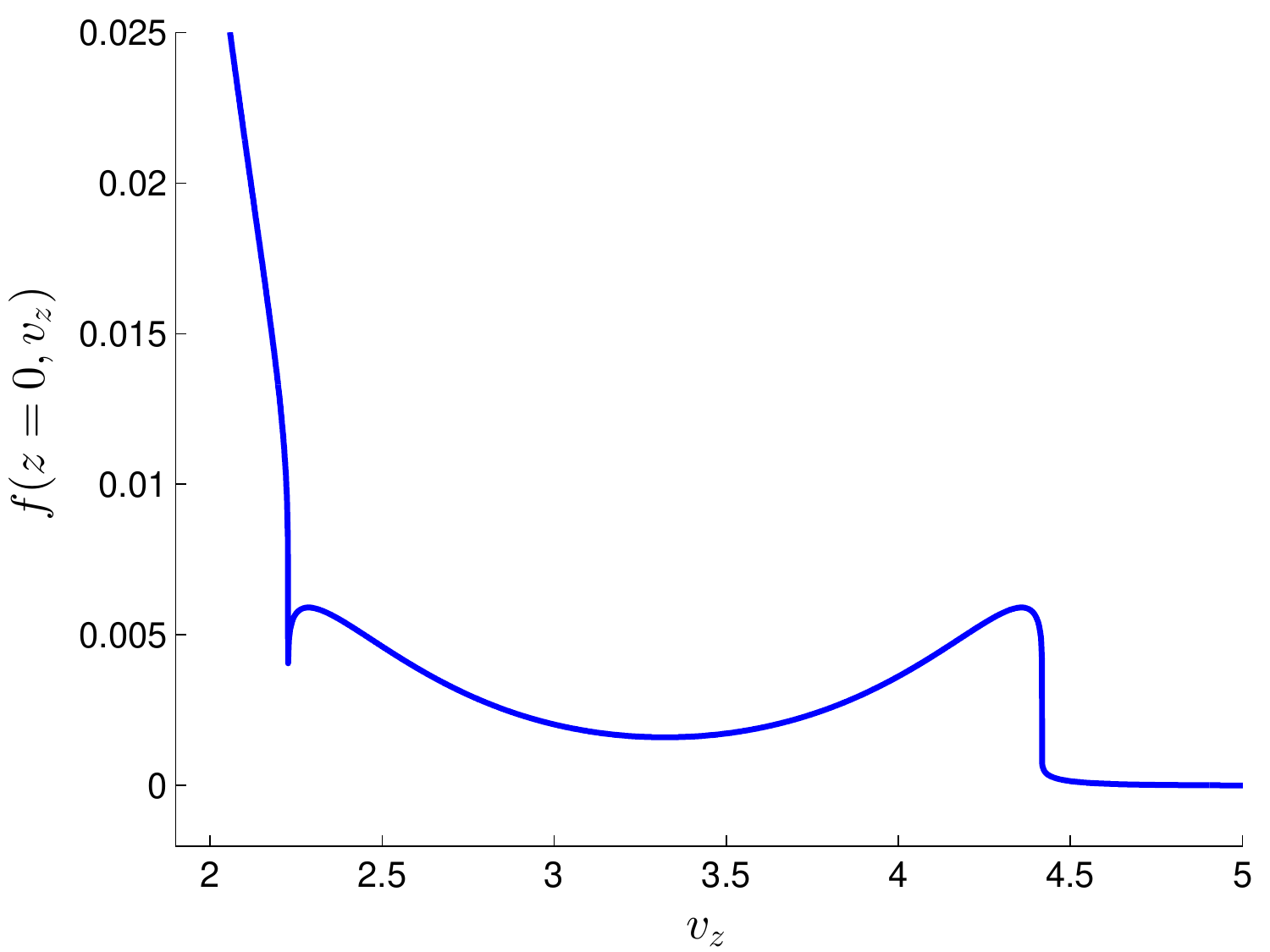}
\caption{Cross-section of $f_{BGK}(z=0,v_z)$ for BGK mode with $k_z=0.35, \phi_0=0.3, v_\varphi=3.321836\ldots$.}
\label{fig:BGKCross}
\end{figure}

Integrating $f_{BGK}$ over $v_z$ that was obtained from Eqs. (\ref{Twint})-(\ref{vdef}) and using the Poisson''s equation \e{eq:Poisson1D}, we get the corresponding electrostatic potential $\Phi_{approx}(z)$ of the approximate BGK solution. One can use $k_z$ as the free parameter  to rescale the solution in such a way that the amplitude of the first harmonic in the electrostatic potential is equal to $\phi_0$ as was assumed in Eq. \e{eq:Phi}.

The result is however only approximate because of higher order Fourier harmonics beyond the fundamental one assumed for $\Phi(z)$ in Eq. \e{eq:Phi}.
   Our calculations show that the second harmonic in $\Phi(z)$ is typically 2-3 orders of magnitude less compared to the first one even for $\phi_0$ of order 1, which validates our initial assumption. We found it satisfactory for the purpose of the subsequent results of this paper  to stop the process of BGK construction at this point. However we also used  $\Phi_{approx}(z)$  to obtain the corresponding  updated $f_{BGK}(z,v_z) $ from Eqs. \e{Twint}-\e{vdef}, calculated second iteration of   $\Phi_{approx}(z)$, and so on. We found that typically  $\sim20$ iterations is sufficient to  converge  $\Phi_{approx}(z)$   within $10^{-15}$ relative   pointwise error over $z$ (with the relative error being $\sim1\%$ after first iteration) to the exact BGK mode.
In this way one can construct a BGK mode for given values of  $\phi_0$ and $v_{\varphi}$ as the input parameters producing the value $k_z$ as the output parameter together with $f_{BGK}.$ If one needs to find  $f_{BGK}$  with the specified  value of  $k_z=k_{z,input}$  then  Newton iterations are performed to find a root of  $k_{z,input} - k_{z,output}(\phi_0,v_{\varphi})=0$  as a function of  either $\phi_0$ or $v_{\varphi}$ keeping the other variable fixed. Here $k_{z,output}(\phi_0,v_{\varphi})$ is the value of $k_z$ obtained for given   $\phi_0$ and $v_{\varphi}$  from the procedure described above.

An example of BGK mode constructed using this approach with Newton iterations over $v_\varphi$  for $k_{z,input}=k_z = 0.35,  \ \phi_0=0.3$ and resulting  $v_\varphi=3.321836\ldots$  is shown in Figs. \ref{fig:BGK}-\ref{fig:BGKCross}. Fig. \ref{fig:BGK} shows  $f_{BGK}(z,v_z)$ around the trapping region with a separatrix $\Phi(z)+v_z^2/2=W=\Phi_{max}$. Fig. \ref{fig:BGKCross} shows the widest cross-section of the the trapping region at $z=0$.

\subsection{BGK dispersion relation and nonlinear frequency shift}
\label{sec:DispersionRelation}
The dispersion relation of the particular family of BGK modes in question has been presented in Refs. \cite{RoseRussellPOP2001} and \cite{RosePOP2005}.
Unlike the linear regime, in which the parameters $k_z$ and $v_\varphi$ are related via well-known $\phi_{eq}$-independent
dispersion relations \cite{VlasovSovPhysUsp1968,Landau1946JPhysUSSR}, a BGK mode's dispersion relation is amplitude dependent. The
BGK mode identified by Eqs. (\ref{eq:1Dmodel}) and (\ref{eq:DoubleLim}) is undamped and has a nonlinear dispersion relation
determined \cite{HollowayDorningPhysRevA1991,BuchananDorningPhysRevE1995} by setting the real part of the dielectric function, $\varepsilon$, to zero. Recall that we define the nonlinear dielectric function $\varepsilon$ as
$\Phi=\Phi_{ext}/\varepsilon$ with $\Phi_{ext}$ being the external pump from SRS, Eq. \e{eq:EXTPotential}. To lowest order  in $\sqrt{\phi_{eq}}$ using $f_{BGK}$ given by Eqs.  \e{Twint}-\e{eq:Phi} one obtains \cite{RoseRussellPOP2001} that,
\begin{equation}
    0=\Real[\varepsilon] \approx \Real[\varepsilon_0] + 1.76f_0''(v_\varphi)\sqrt{\phi_{eq}}/k_z^2,
    \label{eq:Dispersion}
\end{equation}
where
\begin{equation}
    \varepsilon_0(k_z,\omega)=1 - \Xi_0(v_\varphi)/k_z^2,
    \label{eq:DispersionLIN}
\end{equation}
\begin{eqnarray*}
    &\Xi_0(v) = Z'(v/\sqrt{2})/2, \\
    &Z(v) = e^{-v^2}\sqrt{\pi}(i-\erfi(v))=e^{-v^2}(i\sqrt{\pi}-2\int^v_0{e^{t^2}dt}).
    \label{eq:Xi}
\end{eqnarray*}
$Z$ is the plasma dispersion function \cite{FriedGell-MannJacksonWyldJNF1960} and $\varepsilon_0$ is the linear dielectric function.

\begin{figure}
\includegraphics[width=3in]{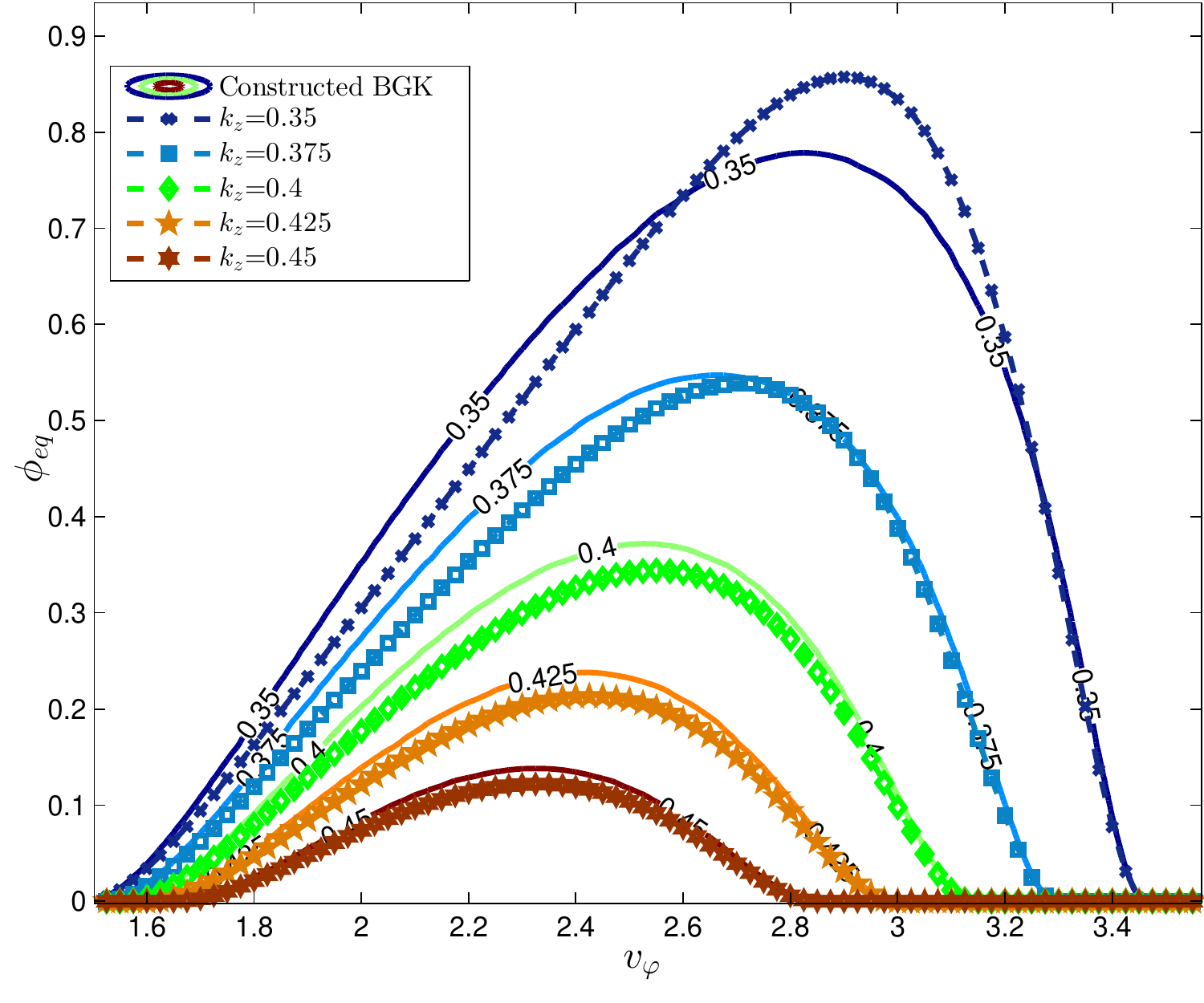}
\caption{(Color online) LW and EAW potential amplitude versus phase velocity for various $k_z$. Solid lines represent constructed BGK family dispersion relation, dashed - approximation of dispersion relation by formula \e{eq:Dispersion}.}
\label{fig:Dispersion}
\end{figure}

Eq. (\ref{eq:Dispersion}) can be solved for $\phi_{eq}(k_z,v_\varphi)$. These solutions are illustrated in Fig. \ref{fig:Dispersion} by dashed lines with markers for various values of $k_z$ together with solid lines corresponding to the BGK modes that were constructed numerically following the procedure in \ref{sec:1DEquilibrium} with the same values of $k_z$ and $v_\varphi$. For $k_z = 0.35$ the maximum amplitude of the constructed BGK is $\phi_{eq} \approx 0.78$
at $v_\varphi \approx 2.85$ while Eq. (\ref{eq:Dispersion}) overestimates the maximum $\phi_{eq}$ at 0.85. The correspondence of solutions of Eq. (\ref{eq:Dispersion}) and values of  $v_\varphi$ for the constructed BGKs for small $\phi_{eq}$ is quite good.  For each $k_z$ and $\phi_{eq}$ less than the maximum amplitude, we have two solutions for $v_\varphi$, the larger value corresponding to the nonlinear LW wave and the smaller one corresponding to the electron acoustic \cite{MontgomeryFociaRoseRussellPRL2001} wave (EAW), similar to two solutions of the Vlasov dispersion relation $\Real[\varepsilon_0(k_z,\omega)]=0$ for a given $k_z$ (see Fig. 2 in Ref. \cite{LancellottiDorningPhysRevLett1998}).

Alternatively, $v_\varphi$ may be considered as a function of $k_z$ and $\phi_{eq}$, i.e. $v_\varphi(k_z,\phi_{eq})$, by
inverting the graph shown in Fig. \ref{fig:Dispersion}. Since a travelling wave's angular frequency, $\omega$, is
always the product of wavenumber and phase velocity, $\omega=k_zv_\varphi$, one may re-express the nonlinear
dispersion relation as a wavenumber and amplitude dependent $\omega$,
\begin{equation}
    \omega(k_z,\phi_{eq}) =k_z v_\varphi(k_z,\phi_{eq}) .
    \label{eq:OMEGA}
\end{equation}

We define the nonlinear frequency shift as
\begin{equation} \label{eq:NLshift}
\Delta\omega^{BGK}=\omega(k_z,\phi_{eq})-\omega_0,
\end{equation}
where $\omega_0=\omega(k_z,\phi_{eq}=0)$. For $k_z=0.35$, $\omega_0=1.21167$.

Expanding $\Real[\varepsilon_0(k_z,\omega)]$ in Eq. \e{eq:Dispersion} in a Taylor series at $\omega=\omega_0$, taking into account that $\Real[\varepsilon_0(k_z,\omega_0)]=0$,  we get an approximation of $\Delta\omega^{BGK}$ given by
\begin{eqnarray}
\label{eq:dW_Rose}
  \Delta \omega_{NL}^{Rose} = -1.76 \left[ \frac{ \p\Real[\varepsilon_0(\omega_0)] }{\p\omega} \right]^{-1} f_0''(v_\varphi) \frac{ \sqrt{\phi_{eq}}}{k_z^2},
\end{eqnarray}
as presented in Eq. (50) of Ref. \cite{RoseRussellPOP2001} and Eq. (9) and Fig. 5 of Ref. \cite{RosePOP2005}. For $k_z=0.35$, $\frac{ \p\Real[\varepsilon_0(\omega_0)] }{\p\omega}=2.335$.

In earlier works of Morales and O'Neil \cite{MoralesNeilPRL1972} and Dewar \cite{DewarPhysFL1972} an approximation for the nonlinear frequency shift of large-amplitude EPW was derived
\begin{eqnarray}
\label{eq:dW_Dewar}
  \Delta \omega_{NL}^{Dewar} = -\alpha \left[ \frac{ \p\Real[\varepsilon_0(\omega_0)] }{\p\omega} \right]^{-1} f_0''(v_\varphi) \frac{ \sqrt{\phi_{eq}}}{k_z^2},
\end{eqnarray}
where $\alpha = 0.77\sqrt{2}=1.089$ and $\alpha=1.163\sqrt{2}=1.645$ for the ``adiabatic" and ``sudden" excitation of nonlinear LW, respectively. The derivation was also summarized in Ref. \cite{BergerBrunnerChapmanPOP2013} and used in Ref. \cite{BergerBrunnerBanksCohenWinjumPOP2015}. In Ref. \cite{RoseRussellPOP2001} after Eq. (48) H. Rose discusses the source of the discrepancy between 1.76 coefficient in Eq. \e{eq:dW_Rose} and 1.645 in Eq. \e{eq:dW_Dewar}.

\subsection{Trapped electron filamentation instability}
\label{sec:TransverseInstability}

LW filamentation instability theory has been presented in Refs. \cite{RosePOP2005} and \cite{BergerBrunnerBanksCohenWinjumPOP2015}, but we
believe that a more cogent and general result was obtained in Ref. \cite{RoseYinPOP2008}, which we now
review.

Let $x$ denote a direction perpendicular to the LW propagation direction, the $z$ axis, with wave amplitude $\phi_{eq}$, the maximum value of $\Phi(z)$ over $z$ (in particular case given by Eq. \e{eq:Phi},  $\phi_{eq}=\phi_0$). Near the equilibrium (BGK mode) in the moving frame, let
\begin{equation}
    \Phi =\Real\left\{\exp(i{\bf k}\cdot {\bf r})[\phi_{eq}+ \delta\phi(t)\exp(i{\bf\delta k \cdot r}) ]\right \},
    \label{eq:Anzac}
\end{equation}
where $\bfk$ is parallel to $z$ direction and $\delta \bfk$ is responsible for the transverse perturbations with the amplitude $\delta\phi(t)
$. Let $\delta\phi \sim exp(\gamma t)$. In Ref. \cite{RoseYinPOP2008} it was shown that
\begin{equation}
    (\gamma + \nu_{residual})^2 =-D\left (\phi_{eq} \frac{\partial\omega}{\partial\phi_{eq}} +D\right),
    \label{eq:GrowtRate}
\end{equation}
wherein the generalized diffraction operator, $D$,
\begin{equation}
    2D =\omega(|\textbf{k}+\delta \textbf{k}|,\phi_{eq})+\omega(|\textbf{k}-\delta \textbf{k}|,\phi_{eq})-2\omega(|\textbf{k}|,\phi_{eq})
    \label{eq:D}
\end{equation}
reduces to the diffraction coefficient, $(|{\bf{\delta k}}_\perp|^2/2 |\bfk|)(\p \omega/\p |\bfk|)+(\delta k_z^2/2)\p^2\omega/\p |\bfk|^2$, for small $|\bf{\delta k}|$. When $\delta{\bf k\cdot k} = 0$, Eq. (\ref{eq:D}) simplifies to
\begin{equation}
    D =\omega(|\textbf{k}+\delta \textbf{k}|,\phi_{eq})-\omega(|\textbf{k}|,\phi_{eq}),
    \label{eq:D2}
\end{equation}
and the (possible) instability is customarily called filamentation, our main regime of interest.

Also assuming  $\phi_{eq} \ll 1$ in  addition to  $\delta{\bf k\cdot k} = 0 $ and  $|\delta{\bf k}|\ll 1$,  we can
 approximate Eq. \e{eq:D2} as follows
\begin{align} \label{eq:Dlin}
    D \approx D_{lin}=\left .\frac{1}{2k_z} \frac{\p \omega(|\textbf{k}|,0) }{\p |\textbf{k}|} \right|_{|\textbf{k}|=k_z}|\delta\textbf{k}|^2\nonumber \\=\frac{v_g}{2k_z} |\delta\textbf{k}|^2, \quad  v_g\equiv\p \omega(|\textbf{k}|,0)/\p |\textbf{k}|,
\end{align}
where $v_g  $ is the group velocity corresponding to the dispersion relation \e{eq:Dispersion} at $\phi_{eq}=0$, $v_g=1.008$ for $k_z=0.35$.

The residual damping, $\nu_{residual}$, from Eq. (\ref{eq:GrowtRate}) is model dependent. For example, if the double limit of Eq. (\ref{eq:DoubleLim}) stops
short of zero value, but with
\begin{equation}
    \frac{\nu_{residual}}{\omega_{bounce}}\ll 1, \ \frac{\phi_{pump}}{\phi_{eq}}\ll1, \ \frac{\omega_{bounce}}{\omega_{pe}} = k_z \sqrt{\phi_{eq}}  ,
    \label{eq:Conditions}
\end{equation}
or in dimensional units, $\frac{\omega_{bounce}}{\omega_{pe}} = k_z\lambda_D \sqrt{e\phi_{eq}/k_BT_e}$.  Then it follows from Eqs. (28), (71) of Ref. \cite{RoseRussellPOP2001} and Fig. 3 of Ref. \cite{RoseRussellPOP2001} that, for $v_\varphi \gtrsim 2.2$,
in dimensional units,\begin{equation}
    \nu_{residual}\thickapprox \frac{\omega_{pe}}{2} \Imag[\varepsilon] \gtrsim \frac{\nu_{SideLoss}}{\omega_{bounce}} \nu_{Landau}.
    \label{eq:NUresidual}
\end{equation}
In Eq. (\ref{eq:NUresidual}), the Landau damping rate, $\nu_{Landau}$, is evaluated for a linear LW with
wavenumber $k_z$.
Also  if $|\bf{\delta k}| \ll |k|$ is not satisfied, it has been argued \cite{BergerBrunnerBanksCohenWinjumPOP2015} that $\nu_{residual}$ is augmented by an
amplitude dependent, but $\nu_{SideLoss}$ independent, form of Landau damping.
However, as we discuss in Section \ref{sec:2DHarveyBGK} below, that addition to $\nu_{residual}$ is not consistent with our simulation results and we set $\nu_{residual}=0$.

\begin{figure}
\includegraphics[width=2.9in]{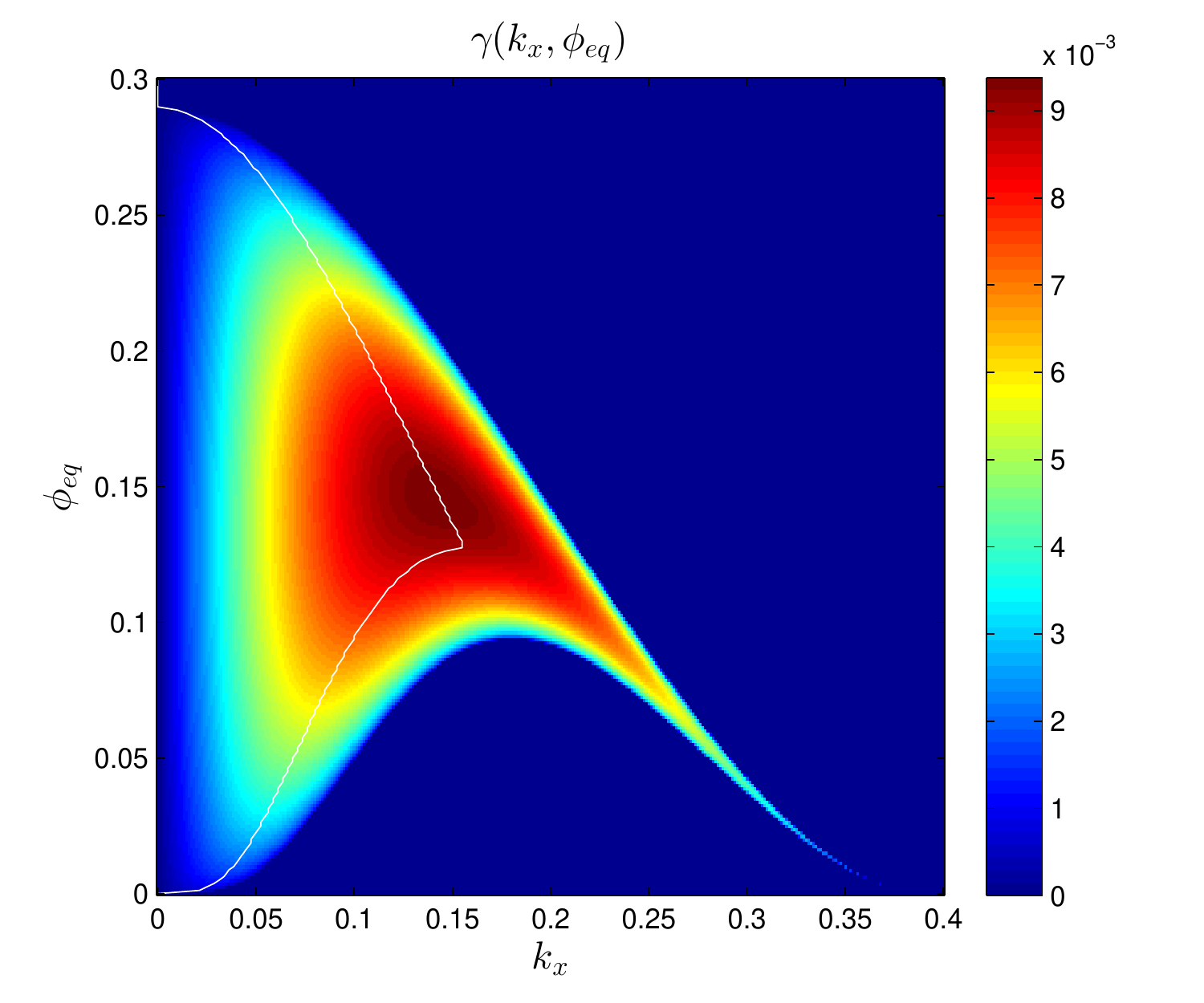}
\caption{(Color online) LW filamentation growth rate contours for $k_z=0.35$. White line shows the maximum growth rate for given $\phi_{eq}$.}
\label{fig:Gkz04}
\end{figure}

\begin{figure}
\includegraphics[width=2.9in]{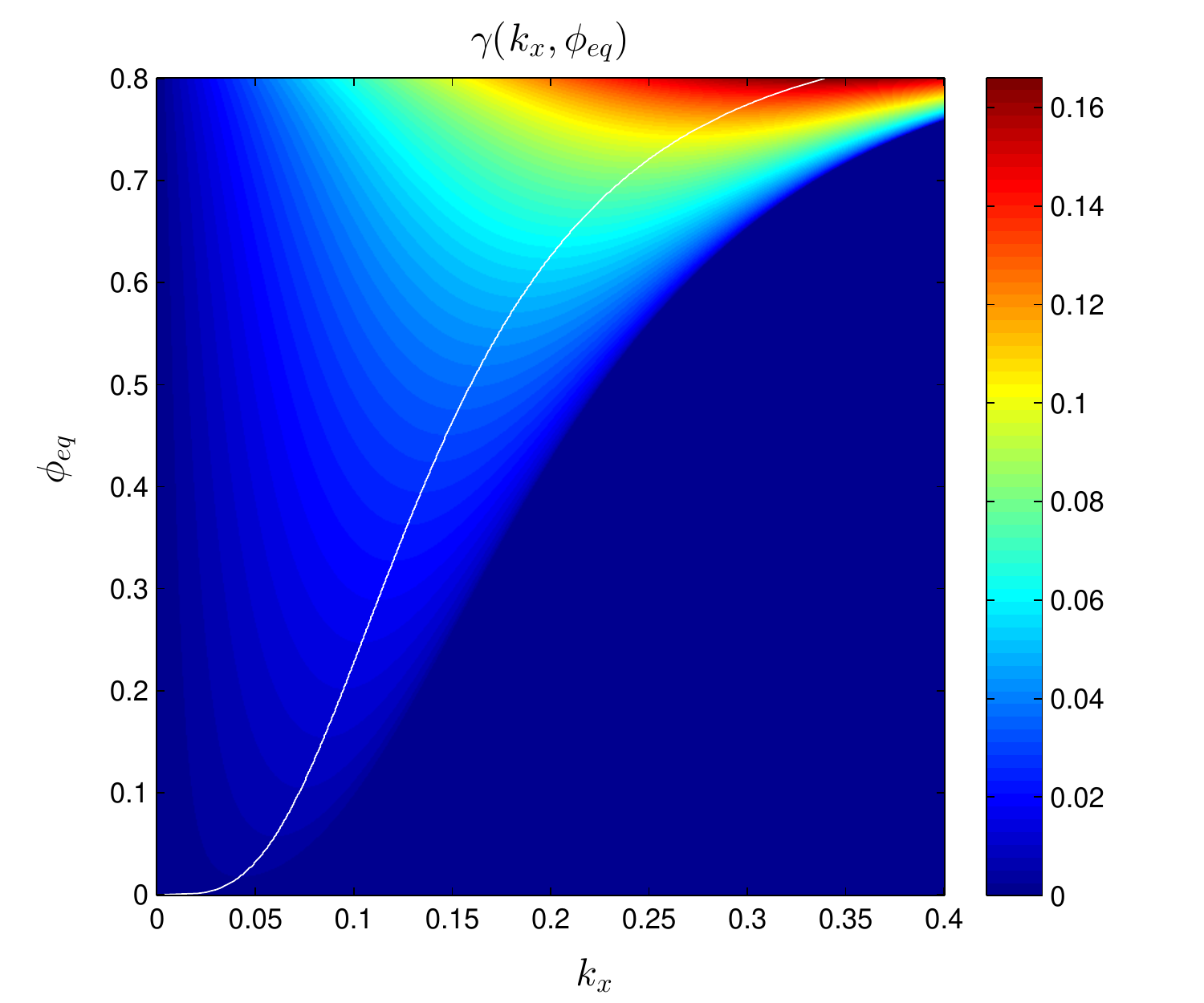}
\caption{(Color online) LW filamentation growth rate contours for $k_z=0.35$ using linear approximation for $D$ as in \e{eq:Dlin}. White line shows the maximum growth rate for given $\phi_{eq}$.}
\label{fig:Gkz035Dlin}
\end{figure}

Eq. (\ref{eq:OMEGA}) may be used to find $\partial\omega/\partial\phi_{eq}$ in terms of $\partial v_\varphi/\partial\phi_{eq}$ which in turn may be
obtained by applying $\partial/\partial\phi_{eq}$ to Eq. (\ref{eq:Dispersion}).
\begin{equation}
    \{Re[\Xi_0'(v_\varphi)] - 1.76f_0'''\sqrt{\phi_{eq}}\}\frac{\p v_\varphi}{\p \phi_{eq}}=\frac{1.76f_0''(v_\varphi)}{2\sqrt{\phi_{eq}}}.
    \label{eq:DerDispersion}
\end{equation}
In the kinetic regime, $D$ may assume negative values as $|\bf{\delta k}|$ \cite{RosePOP2005} and/or $\phi_{eq}$
\cite{RoseYinPOP2008} increase. Therefore, the qualitative shape of $\gamma$ contours determined by Eqs. (\ref{eq:Dispersion}), (\ref{eq:GrowtRate}) and (\ref{eq:D}) may
differ from fluid model modulational/filamentation \cite{BergerBrunnerBanksCohenWinjumPOP2015}, whose domain of
applicability is limited, a priori, to $k_z\ll1$.
Contours of $\gamma$ for $k_z = 0.35$ are shown in Figs. \ref{fig:Gkz04}, with $\nu_{residual}$ set to $0$. For more contours of $\gamma$ and related figures see \cite{RoseYinPOP2008}.  If we use linear approximation for $D$ as in Eq. (\ref{eq:Dlin}) and solve Eqs. (\ref{eq:Dispersion}), (\ref{eq:GrowtRate}) for $k_z = 0.35$ and  $\nu_{residual}=0$ we get contours of $\gamma$ as shown in Fig. \ref{fig:Gkz035Dlin}. As $k_z$ is increased, the range of amplitudes over which Eqs. (\ref{eq:Dispersion}), (\ref{eq:GrowtRate}) and (\ref{eq:D}) predict growth is reduced, while using $D_{lin}$ from Eq. (\ref{eq:Dlin}) provides growth in a full range of amplitudes for any $k_z$. The latter case in more consistent with the simulations as we will see in Section \ref{sec:NUMERICALSIMULATIONS}.

Another simplification can be made if one assumes at the leading order that the nonlinear frequency shift $\Delta\omega\varpropto\sqrt{\phi_{eq}}$. Then $\phi_{eq} \frac{\partial\omega}{\partial\phi_{eq}}=\Delta\omega/2$ and maximizing $\gamma$ over $D$ in Eq. \e{eq:GrowtRate}
we obtain the maximum value
\begin{equation} \label{gammamax}
\gamma^{max}=|\Delta\omega|/4,
\end{equation}
at
\begin{equation} \label{Dmax}\
D=-\Delta\omega/4,
\end{equation}
which is valid for $|\bf{\delta k}| \ll |k|$ and $\nu_{residual}=0$.
Using the approximation \e{eq:Dlin}, we obtain from Eq. \e{Dmax} the position of the maximum
\begin{equation} \label{kmaxapprox}
|{\bf{\delta k}}| = k_x^{max}=\left ( \frac{-\Delta\omega \,k_z}{2v_g}\right )^{1/2}.
\end{equation}
%

\section{NUMERICAL SIMULATIONS of LW filamentation }
\label{sec:NUMERICALSIMULATIONS}
 Here we describe $2+2D$ fully nonlinear Vlasov simulations that we performed to study the filamentation instability of BGK modes described in the previous section.

\subsection{Simulation settings and methods}
\label{sec:Simulationsettings}

We simulate $2+2D$ Vlasov-Poisson system  (\ref{eq:vlasov})-(\ref{eq:density}) in phase space, $(z,v_z,x,v_x)$,  using fully spectral (in all four dimensions) code and 2nd order in time split-step (operator splitting) method with periodic boundary conditions (BC) in all four dimensions. To ensure spectral convergence and imitate the weak effect of collisions, we added to Eq. (\ref{eq:vlasov})  a small hyper-viscosity term as follows

\begin{equation}
\begin{split}
     &\left\lbrace\frac{\partial }{\partial t} + v_z\frac{\partial }{\partial z} + v_x\frac{\partial }{\partial x} + E_z \frac{\partial }{\partial v_z} + E_x \frac{\partial }{\partial v_x} \right\rbrace f= \\  
     &- D_{16v_z}\frac{\p^{16}}{\p v_z^{16}}\left(f - \frac{1}{L_z}\int_0^{L_z} f dz\right),
    \label{eq:2DVlasov}
\end{split}
\end{equation}
where  $D_{16v_z}$ is the 16th order hyper-viscosity coefficient. We use periodic BC in $z$ direction with period $L_z=2\pi/k_z$ and $k_z=0.35$ in our simulations. Choosing   $L_z=2\pi/k_z $ allows us to focus on the study of filamentation instability effects (along $x$) while avoiding subharmonic (sideband) instability \cite{KruerDawsonSudanPRL1969} in the longitudinal $z$-direction.  Periodic BC in $x$ with the period $L_x$ together with  $x$-independent initial condition (IC) are used to separate filamentation instability effects from any sideloss effects due to trapped electrons traveling in the transverse direction (this is in contrast to Ref. \cite{LushnikovRoseSilantyevVladimirovaPhysPlasmas2014}, where the transverse spatial profile in the initial condition made sideloss comparable with filamentation instability growth rate).  We chose typically $200\pi\le L_x\le 800\pi$ depending on the BGK mode's amplitude to capture all growing transverse modes. Periodic BC in $v_z$ and $v_x$ were used  without sacrificing any accuracy of the simulation compared to outgoing BC since the particle flow through the boundary at $v_z=v_z^{max}$ is  $\propto E_z\frac{\p f}{\p v_z}$ with $\frac{\p f}{\p v_z}\approx \frac{v_z}{\sqrt{2\pi}} e^{-\frac{v_z^2}{2}}$ which  can be made as small as desired by picking large enough $v_z^{max}$. Typically we choose $v_z^{max}=8$ for which $|E_z\frac{\p f}{\p v_z}|\approx 10^{-15}$. The same argument is applied in $v_x$ direction with the only difference that in our simulations $E_x$  is several orders less than $E_z$ so $v_x^{max}$ can be chosen smaller than $v_z^{max}$. Typically we choose $v_x^{max}=6$ for which the flow through $v_x=v_x^{max}$ boundary is at the level of machine precision.

Split-step method of 2nd order was chosen over other methods since it is unconditionally stable (which allows large time steps), preserves number of particles at each time step exactly and has a very small error in the full energy of the system. That error is not accumulated over time (in contrast with Runge-Kutta methods where such accumulation occurs). We also decided to choose 2nd order method over higher order methods because our experiments with the size of time step and methods of various orders showed that the time integration error is dominated by the errors coming from other sources (space discretization and hyper-viscosity term).

The hyper-viscosity term in the right-hand side (r.h.s.) of Eq. \e{eq:2DVlasov} is used to prevent recurrence \cite{ChengKnorrJCompPhys1976} and aliasing (which causes propagation of numerical error from high modes to low modes) effects. The hyper-viscosity operator in r.h.s. of Eq. \e{eq:2DVlasov} has to be a smooth function in the Fourier transformed $v_z$ space. At the same time we found it beneficial to use high-order (here we choose 16th order) over low-order hyper-viscosity since it affects low modes of solution less while having effectively the same damping effect on high modes.  That allows to use a smaller numerical grid for the same overall precision. The coefficient $D_{16v_z}$ is chosen as small as possible to prevent aliasing depending on the resolution of simulation in $v_z$ directions. Our safe estimate $D_{16v_z}\approx |\gamma_{Landau}(k_z)| (\frac{2\Delta v_z}{\pi})^{16}$ with $\gamma_{Landau}(k_z=0.35)=-0.034318\ldots$ found to be sufficient to avoid aliasing issues and completely remove the recurrence effect \cite{ChengKnorrJCompPhys1976} in linear Landau damping simulations (while still recovering proper Landau damping with any desired accuracy for simulations with low-amplitude waves). Simulations with high amplitude waves (with $\Phi\sim1$) might require higher value of hyper-viscosity coefficient $D_{16v_z}$, so one needs to keep track of spectrum of the solution in $(z,v_z)$ space and adjust $D_{16v_z}$ if needed. We typically used $D_{16v_z}=10^{-25}$ for simulations with $N_z\times N_{v_z}=64\times  256$ grid points in $(z,v_z)$ space and $D_{16v_z}=10^{-30}$ for $N_z\times N_{v_z}=128\times  512$. Also hyper-viscosity does not affect conservation of number of particles in the system while having positive effect on conservation of energy in long-term simulations. While the term $-\frac{1}{L_z}\int_0^{L_z} f dz$  in r.h.s. side of \e{eq:2DVlasov} is not absolutely necessary, we found that the total energy of the system is conserved better if this term is used. This is because this term  prevents filtering out of the 0th harmonic of $f$ in $z$-space that holds most of the kinetic energy. We did not need any hyper-viscosity in $v_x$ direction since the electrostatic field (and therefore both perturbations of electron density and amount of energy in high modes) in transverse direction is many orders of magnitude weaker compared to the longitudinal direction $(z,v_z)$ throughout most of the simulation until nonlinear self-focusing event at the end. Detailed simulation of that event is however  outside of the scope of this paper.

All simulations are carried out in the lab frame rather then in moving frame, since in this case the tails of the distribution function in $v_z$ direction are almost symmetric and have smaller values $\propto \exp(-(v_z^{max})^2/2) $ at the boundaries $\pm v_z^{max}$ compared to the tail value $\propto \exp(-(v_z^{max}-v_\varphi)^2/2) $ in simulations done in the wave frame moving with velocity $v_\varphi$ with the same $v_z^{max}$. For this reason  simulations performed in the lab frame have smaller numerical error due to periodic BC in $v_z$.

\subsection{2+2D simulations and filamentation instability of  1D BGK modes}
\label{sec:2DHarveyBGK}
\begin{figure}
\includegraphics[width=2.8in]{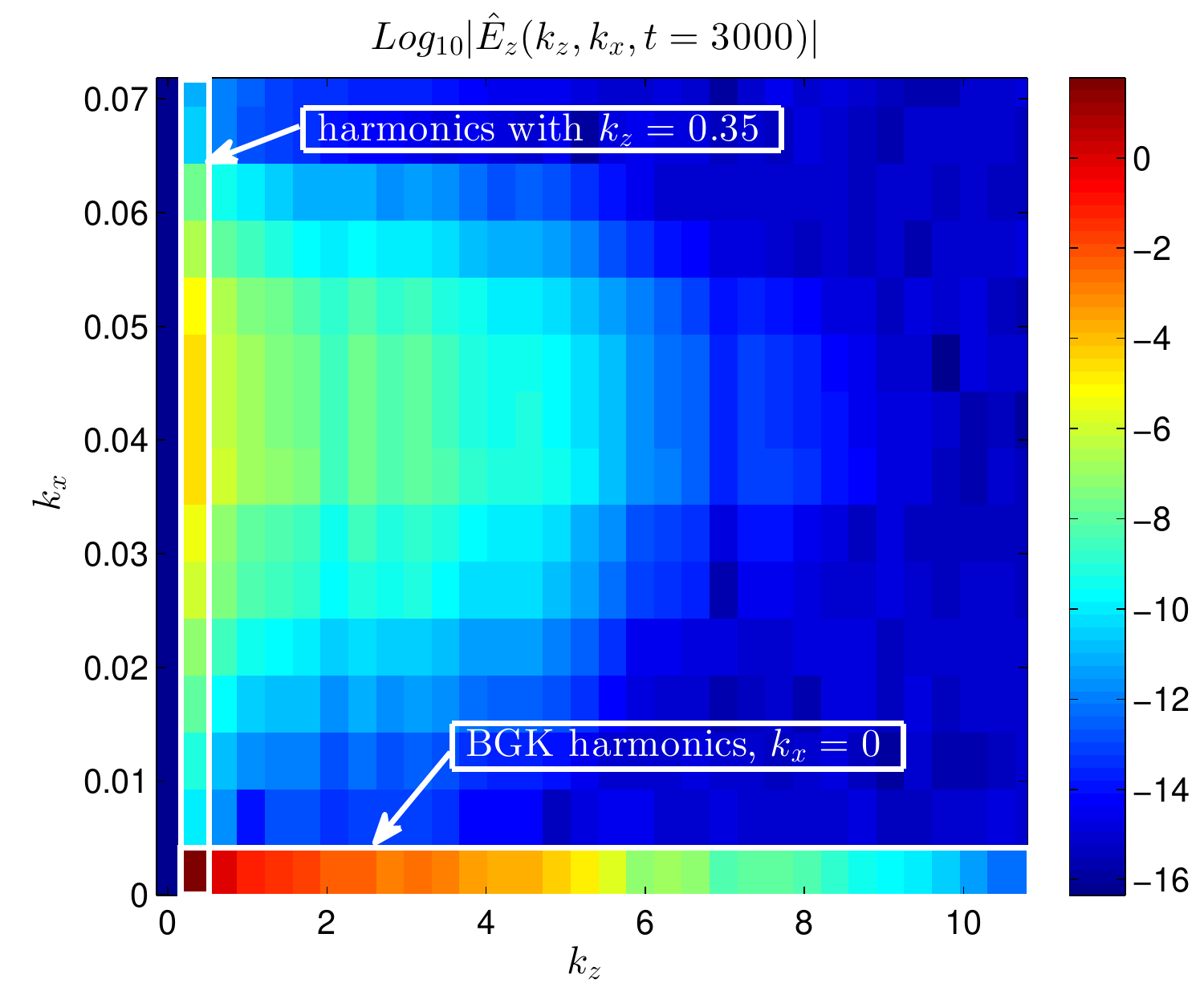}
\caption{(Color online) The density plot of the spectrum of $E_z(z,x)$ at  $t=3000$.}
\label{fig:EzSpectrum}
\end{figure}

\begin{figure}
\includegraphics[width=2.6in]{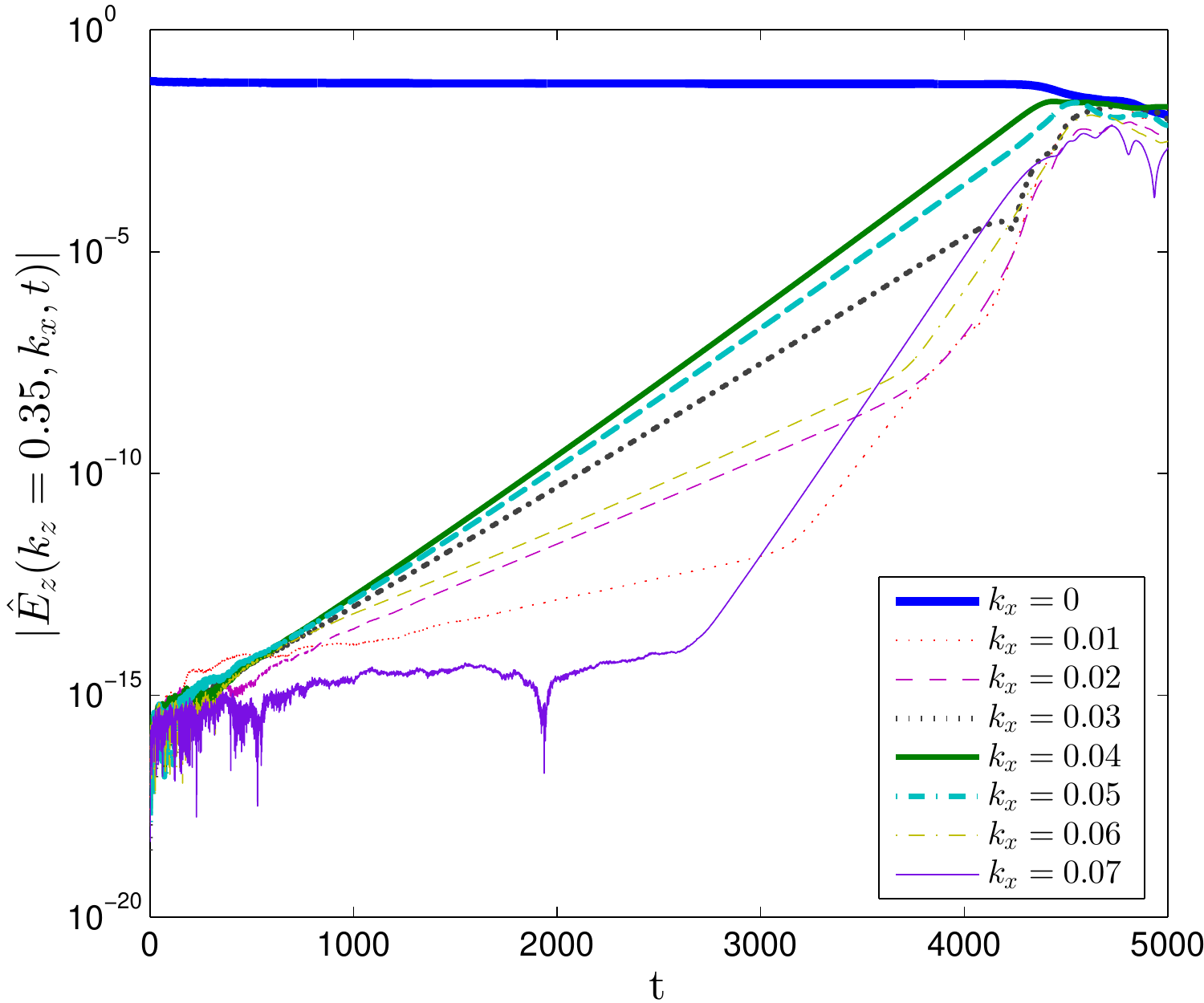}
\caption{(Color online) The growth of harmonics $|\hat E_z(k_z=0.35,k_x,t)|$ in time.}
\label{fig:EzGrowth}
\end{figure}

\begin{figure}
\includegraphics[width=2.8in]{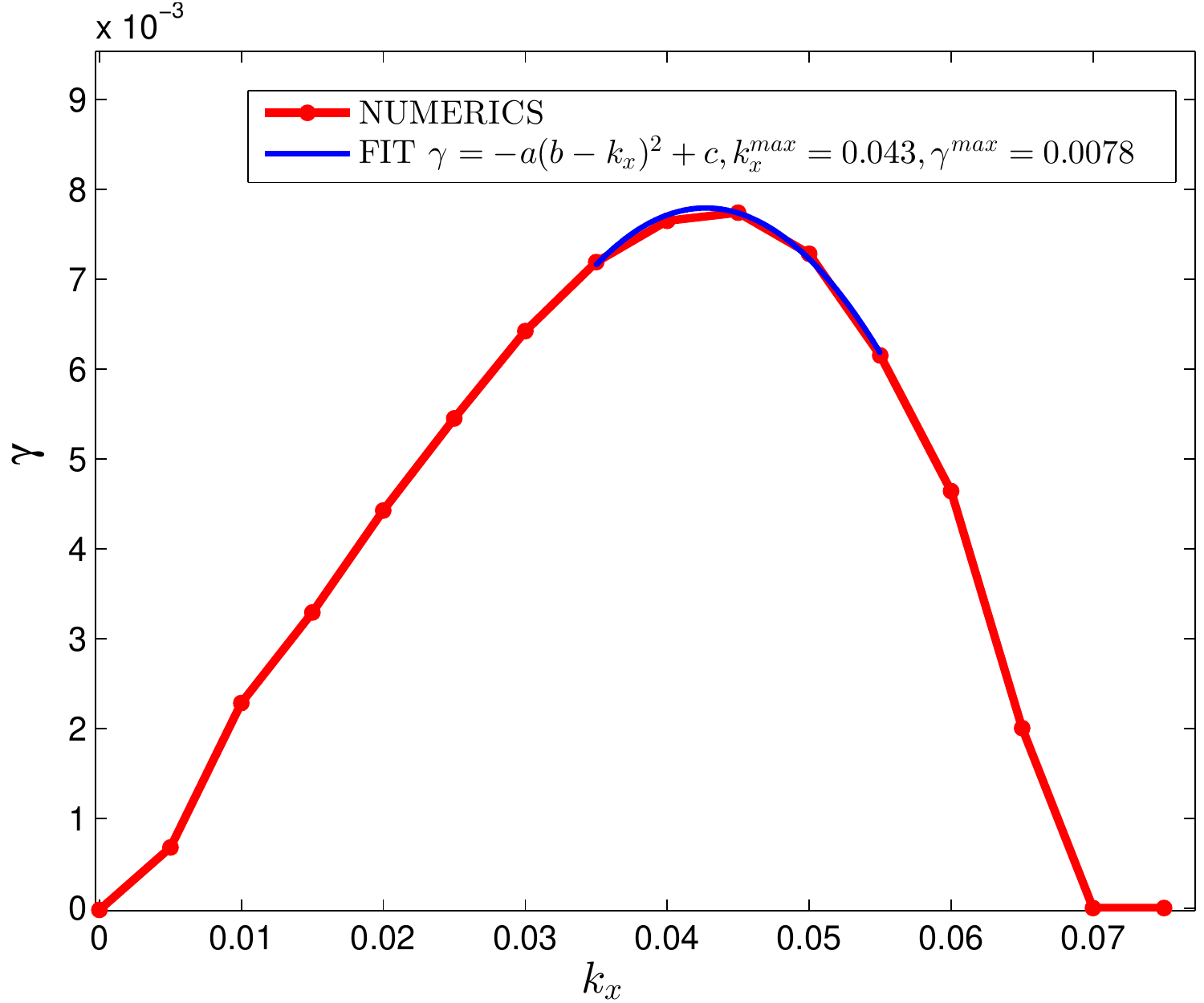}
\caption{(Color online) The growth rates $\gamma_{k_x}$ of oblique harmonics extracted from the least-square fit to the data of Fig. \ref{fig:EzGrowth}. A fit to the quadratic law near the maximum is also shown.}
\label{fig:G(kx)}
\end{figure}

\begin{figure}
\includegraphics[width=2.8in]{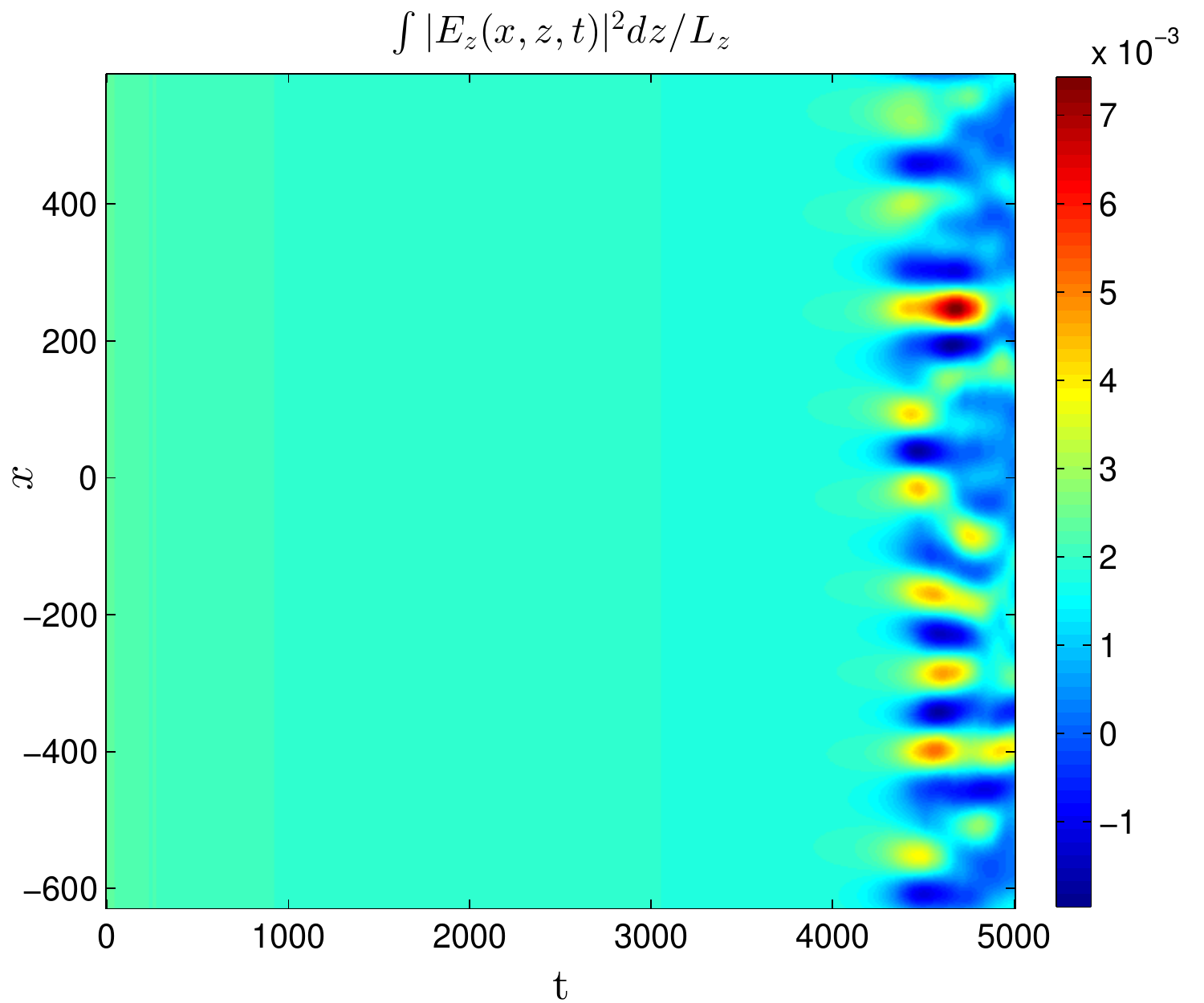}
\caption{(Color online) The density plot vs.  $x$ and $t$ for   $\langle|E_z|^2\rangle_z\equiv L_z^{-1}\int^{L_z}_0|E_z|^2dz$ ($|E_z|^2$
averaged over $z$) shows a development of LW filamentation with time from the initial BGK mode. }
\label{fig:Filamentation}
\end{figure}

\begin{figure}
\includegraphics[width=2.8in]{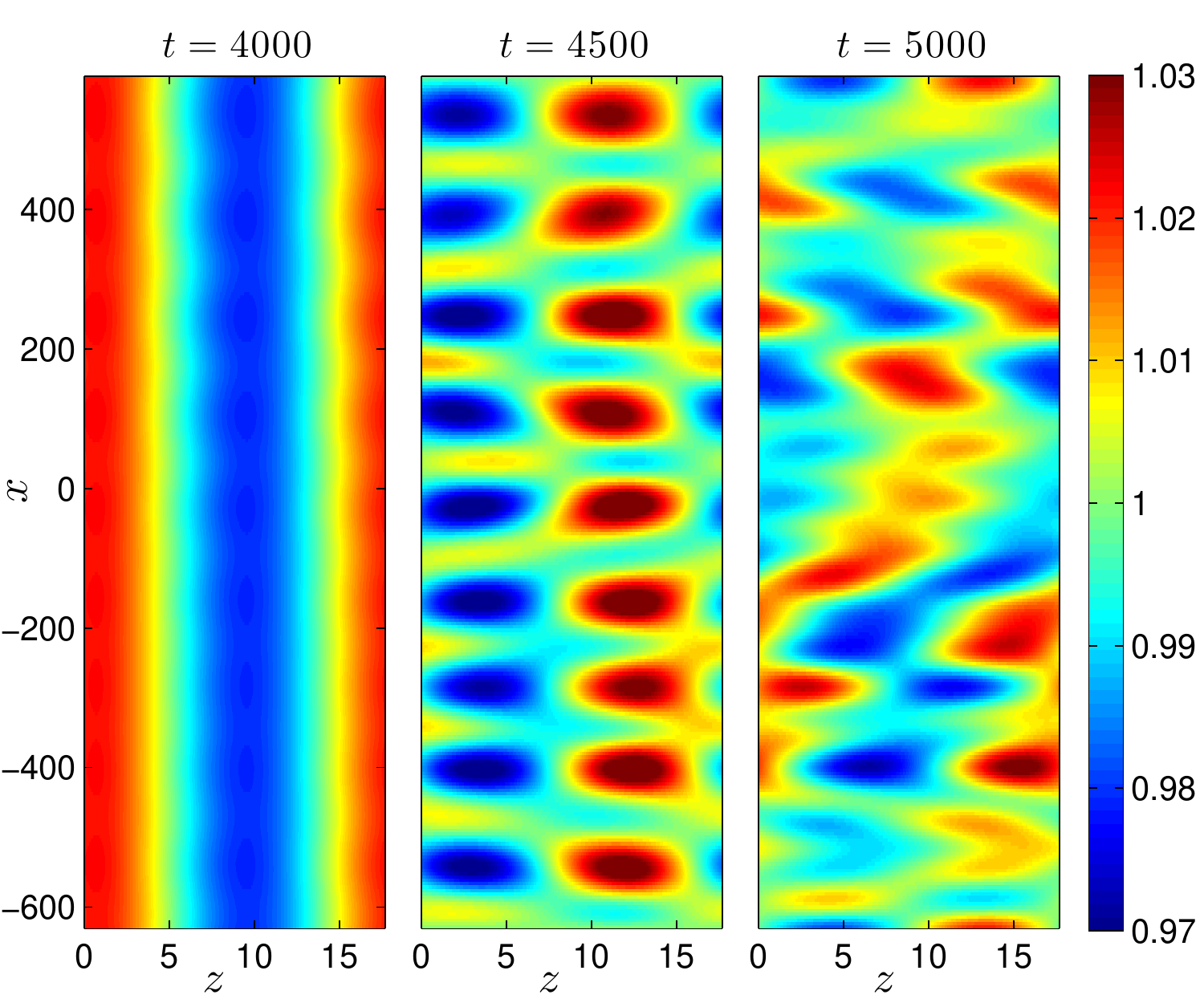}
\caption{(Color online) Modulation of particle density $\rho(z,x)$  before ($t=4000$), during ($t=4500$) and after ($t=5000$) Langmuir wave filamentation.}
\label{fig:RhoModulation}
\end{figure}

\begin{figure}
\includegraphics[width=2.8in]{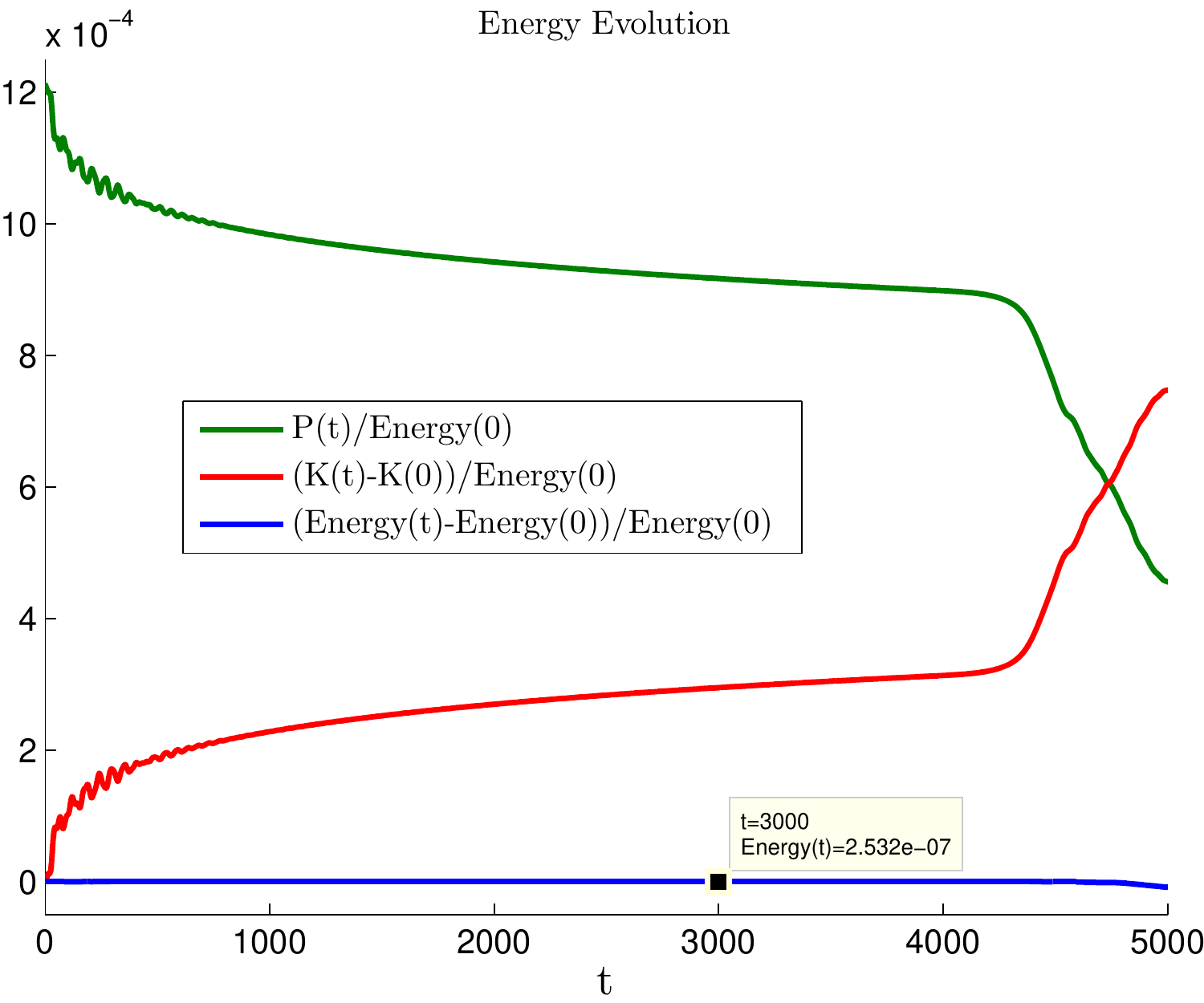}
\caption{(Color online) Evolution of electrostatic, P($t$), kinetic, K($t$), and total energy, Energy($t$), in the simulation with BGK amplitude $\phi_{eq}$=0.2.}
\label{fig:Energy}
\end{figure}

In these simulations we use IC of the form of Eq. \e{eq:BGKMaxwellian} that has the constructed BGK mode from Section \ref{sec:1DEquilibrium} in the $(z,v_z)$ directions, uniform in the $x$-direction and a Maxwellian distribution $f_0(v_x)$ in the $v_x$ direction,
\begin{equation}
    f(z,v_z,x,v_x,t=0) =f_{BGK}(z,v_z)f_0(v_x).
    \label{eq:IC1}
\end{equation}
We run  simulations for a long enough time to observe the growth of oblique harmonics of electric field with wave vectors $(k_z=0.35,k_x)$ (see Fig. \ref{fig:EzSpectrum} for a quarter  of $E_z(z,x)$ spectrum, other quarters of the spectrum are similar to it) for several orders in magnitude  (see Fig. \ref{fig:EzGrowth}), where $k_z$ is the wavenumber corresponding to the  BGK mode and $k_x$ varies between $-k_x^{max}$ and $k_x^{max}=\pi/\Delta x$, $\Delta x=L_x/N_x$, where $N_x$ is the number of grid points in $x.$ The initial values in these harmonics are near the machine precision from the round-off errors. During the simulation they grow from values $\sim10^{-16}$ to $\sim10^{-1}$. The exponential growth rates $\gamma_{k_x}$ for these harmonics are extracted (see Fig. \ref{fig:G(kx)}) from the least-square fit  when the amplitudes grow from $\sim10^{-13}$ to $\sim10^{-8}-10^{-6}$ (during these times a clear exponential growth $\propto e^{\gamma_{k_x}t}$ is observed). Later in the simulation, nonlinear self-focusing effects come into play and LW filamentation occurs (see Figs. \ref{fig:Filamentation} and \ref{fig:RhoModulation}) transferring a significant part of electric field energy, P($t$)=$\iint \frac{|E_z|^2+|E_x|^2}{2}dzdx$, into kinetic energy,
K($t$)=$\iiiint\frac{(v_z^2+v_x^2)}{2}fdzdxdv_zdv_x $ (see Fig. \ref{fig:Energy}). Notice also that the relative error in full energy of the system, Energy($t$)=P($t$)+K($t$), is small. Figs.  \ref{fig:EzSpectrum}-\ref{fig:Energy}  are obtained from the simulation with $\phi_{eq}=0.2$. Other simulation parameters were $D_{16v_z}=10^{-25}$, $64\times 256\times 64\times 32$ grid points for $(z,v_z,x,v_x)$ with $L_z=2\pi/k_z, L_x=400\pi, v_z^{max}=8,v_x^{max}=6$, the time step $\Delta t=0.1$ and the final simulation time $T_{final}=5000$.
Simulations with a larger $L_x$ and correspondingly larger extent of spectrum in $k_x$ were done too but no other regions of growing modes in spectrum (such as in Figs. \ref{fig:EzSpectrum} and \ref{fig:G(kx)}) were observed except for the one starting near $k_x=0$.


These simulations were done for a variety of BGK modes with $k_z=0.35$, amplitudes $0.025 \leq \phi_{eq} \leq 0.77$ and values of $v_\varphi$ according to the BGK dispersion relation \e{eq:Dispersion}. The parameters of these simulation were $D_{16v_z}=10^{-25}, 64\times 256\times 32\times 32 $ grid points for $(z,v_z,x,v_x), \Delta t=0.1$ and $ 2000\le T_{final}\le 30000$ (depending on BGK amplitude). Another set of simulations was performed for $D_{16v_z}=10^{-30}$ and $128\times 512\times 32\times 32$ grid points with the rest of parameters being the same.

\begin{figure}
\includegraphics[width=2.8in]{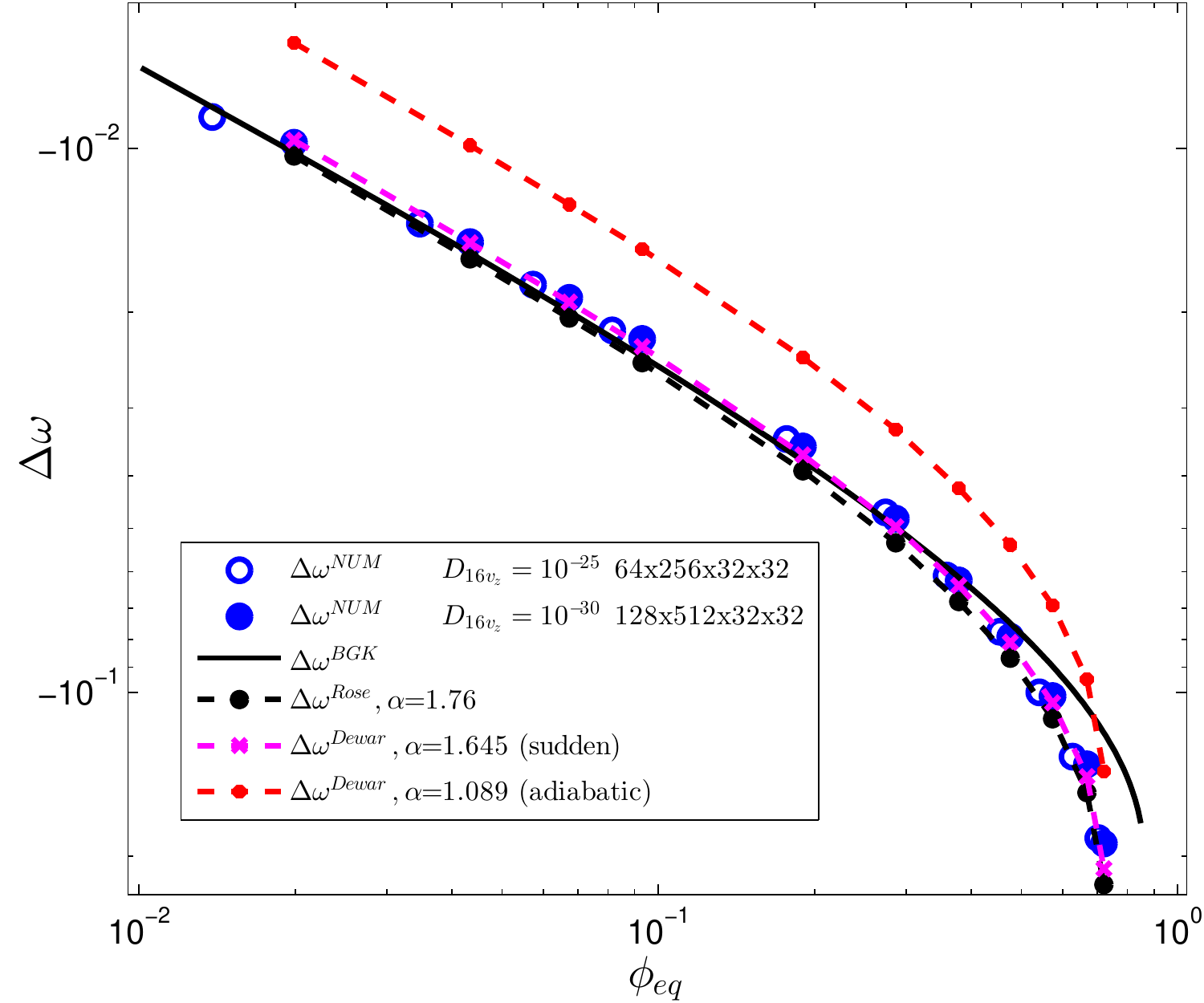}
\caption{(Color online) Nonlinear frequency shift as a function of BGK amplitude $\phi_{eq}$.}
\label{fig:delta_W_vs_phi}
\end{figure}

We extract the nonlinear frequency shift $\Delta\omega^{NUM}$ from simulations by finding the wave frequency as the rate of change of the phase of the Fourier harmonic of  $\Phi$  with $k_z=0.35,$  $k_x=0$ and subtracting the frequency that corresponds to our undamped BGK mode in the limit of zero amplitude, $\omega_0=\omega(k_z=0.35,\phi_{eq}=0)=1.2116687\ldots$, which can be found as a real root of $\Real[\varepsilon_0(k_z,\omega)]=0$ or Eq. \e{eq:Dispersion} with $\phi_{eq}=0$. Note that the frequency of the damped linear LW (real part of a complex root of $\varepsilon_0(k_z,\omega)=0$  \cite{PitaevskiiLifshitzPhysicalKineticsBook1981,NicholsonBook1983}) is $\omega_{LW}(k_z=0.35)=1.22095\ldots$, for the discussion of Vlasov vs. Landau analysis see Ref. \cite{BuchananDorningPhysRevE1995}. The difference is $\approx1\%$ for $k_z=0.35$ and it becomes larger for larger $k_z$. Fig. \ref{fig:delta_W_vs_phi} shows $\Delta\omega^{NUM}$ (large circles (blue color online)) obtained from simulations in comparison with theoretical one $\Delta\omega^{BGK}$ (solid black line) computed using Eq. \e{eq:NLshift}, approximations $\Delta\omega_{NL}^{Rose}$ (dashed black line with circle markers) and $\Delta\omega^{Dewar}$ (dashed grey lines with small ``o" (red online) and ``x"(pink online) markers) given by Eqs. \e{eq:dW_Rose} and \e{eq:dW_Dewar}, respectively, for which we used $v_\varphi=\omega^{NUM}/k_z$.
We conclude from Fig. \ref{fig:delta_W_vs_phi} that both $\Delta\omega_{NL}^{Rose}$ and $\Delta\omega^{Dewar}$ with $\alpha=1.645$ (sudden) work really well for the whole range of amplitudes whereas $\Delta\omega^{BGK}$ works well for amplitudes of BGK $\phi_{eq}\lesssim 0.5$ since $v_\varphi$ for $\phi_{eq}>0.5$ deviates from the solution of approximate dispersion relation Eq. \e{eq:Dispersion} as can be seen in Fig. \ref{fig:Dispersion}.
\begin{figure}
\includegraphics[width=2.8in]{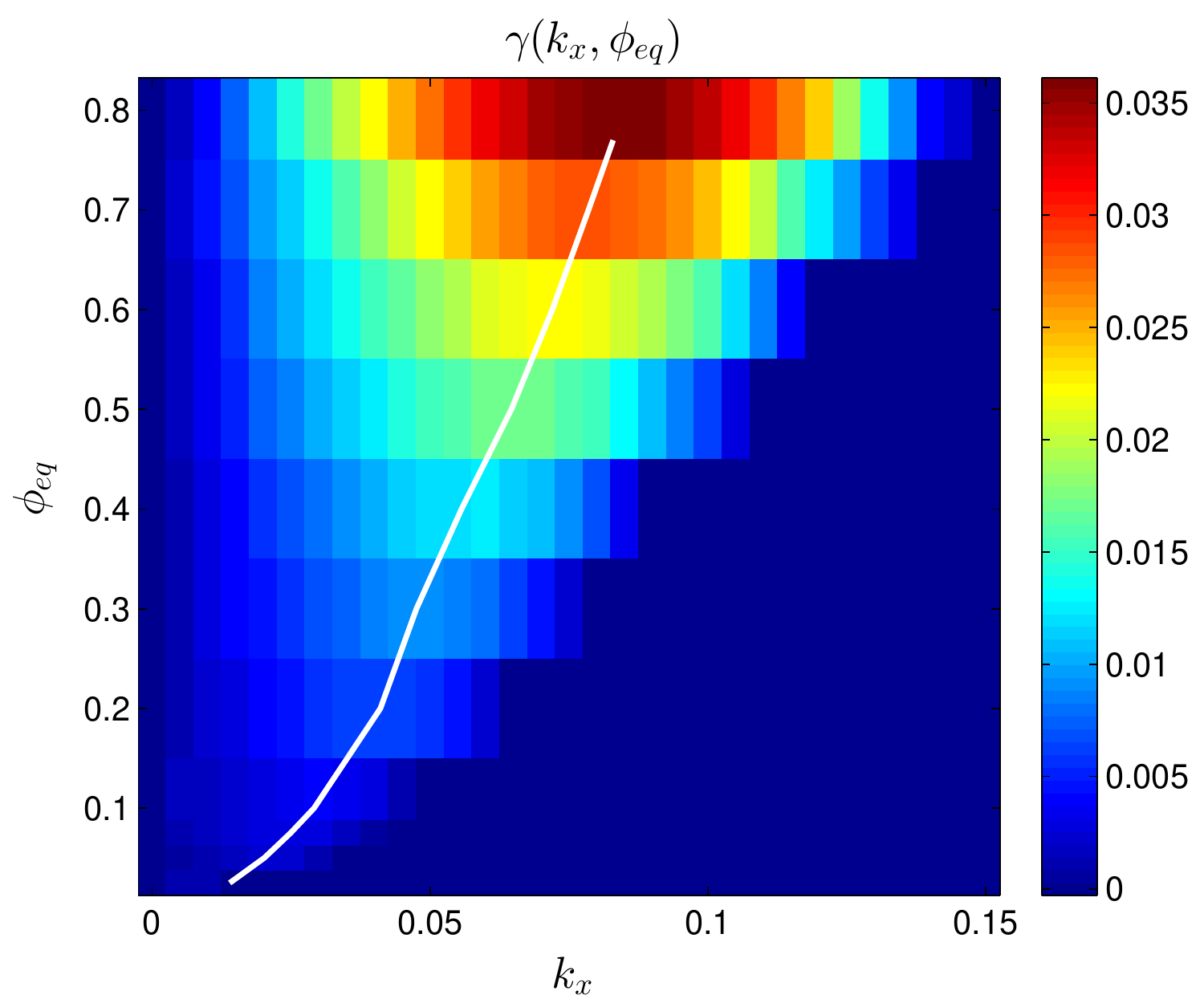}
\caption{(Color online) Numerical growth rates $\gamma$ density plot as a function of $k_x$ and BGK amplitude $\phi_{eq}$. The white line shows the position of the maximum  $\gamma_{k_x}$ for each $\phi_{eq}.$        }
\label{fig:G_vs_kx_vs_phi}
\end{figure}
\begin{figure}
\includegraphics[width=2.6in]{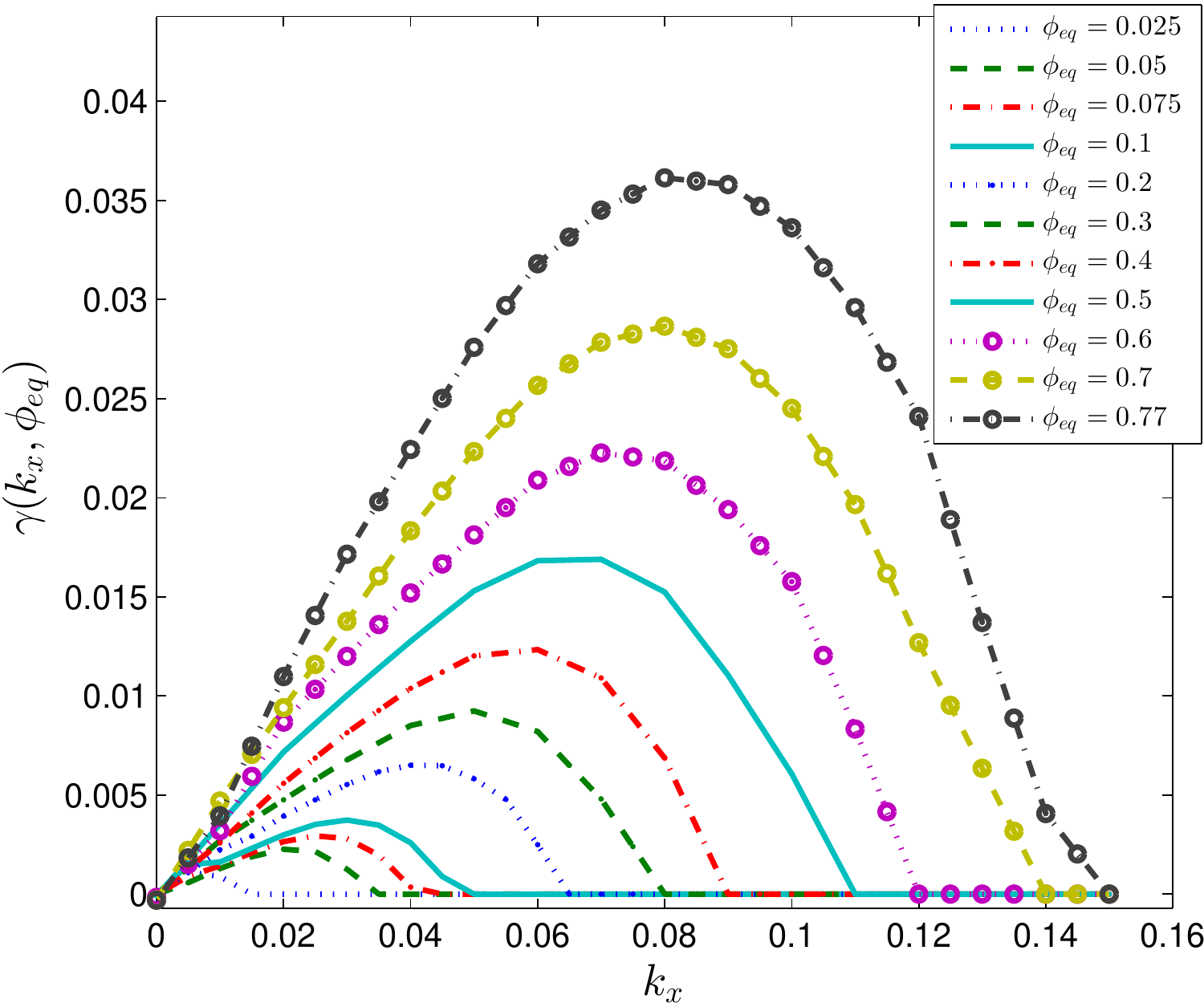}
\caption{(Color online) The growth rates  $\gamma_{k_x}$   as a function of $k_x$ for BGK modes with various amplitudes $\phi_{eq}$ correspond to multiple cross-sections of Fig. \ref{fig:G_vs_kx_vs_phi}.}
\label{fig:G_vs_kx_vs_phi_}
\end{figure}

\begin{figure}
\includegraphics[width=2.8in]{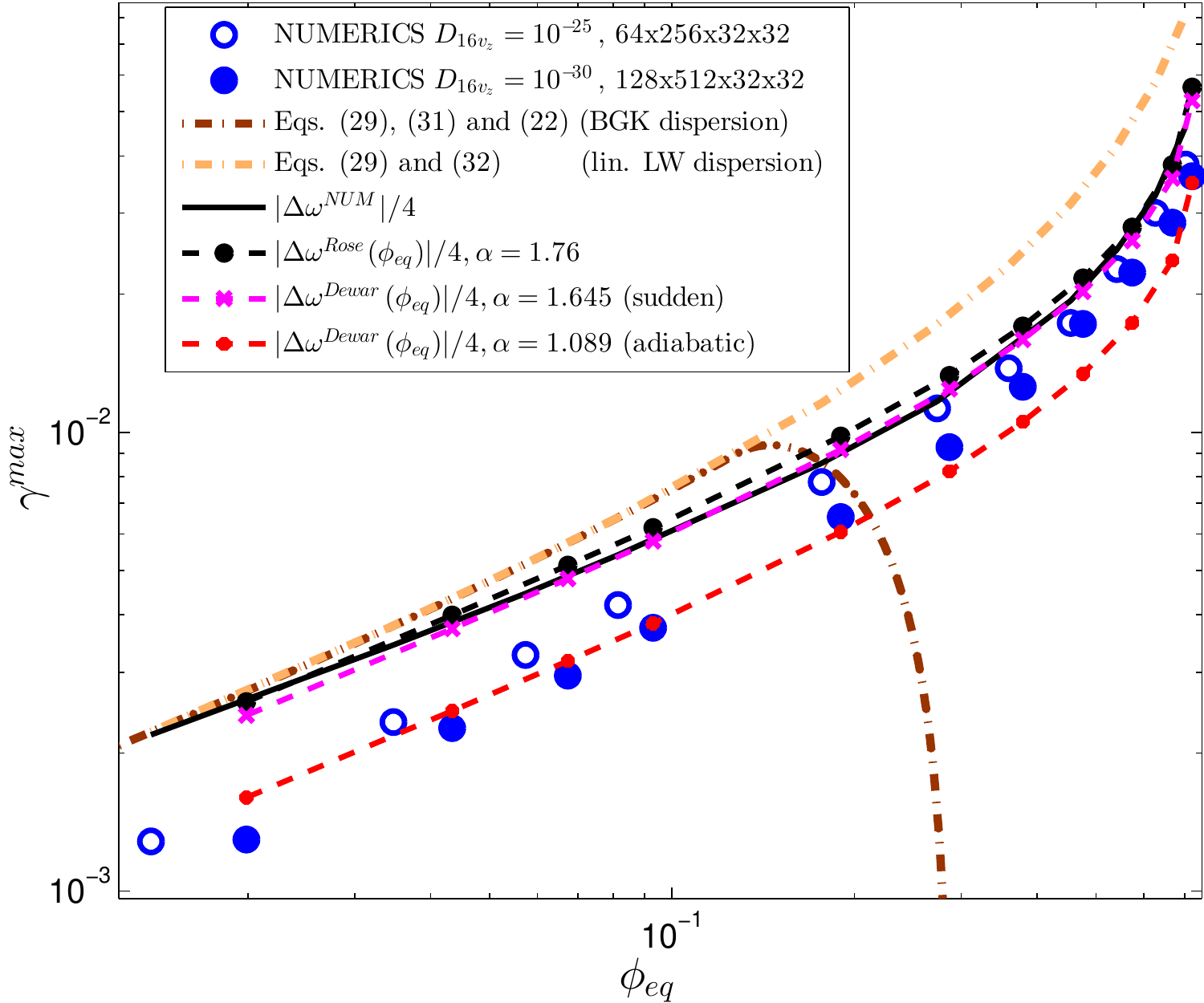}
\caption{(Color online) The maximum growth rate as a function of BGK amplitude $\phi_{eq}$.}
\label{fig:G_max_vs_phi}
\end{figure}

Growth rates of filamentation instability as a function of $k_x$ from the series of simulations with $D_{16v_z}=10^{-30}$ and various amplitudes  $\phi_{eq}$ are given in Figs. \ref{fig:G_vs_kx_vs_phi} and \ref{fig:G_vs_kx_vs_phi_}. The maximum growth rate $\gamma^{max}$ (the maximum vs. $k_x$ for each fixed  $\phi_{eq}$ ) as a function of $\phi_{eq}$ is shown in Fig. \ref{fig:G_max_vs_phi} (for $D_{16v_z}=10^{-30}$ and $D_{16v_z}=10^{-25}$, filled and non-filled circles, respectively) together with the theoretical predictions given by Eqs. \e{eq:Dispersion}, \e{eq:OMEGA}, \e{eq:GrowtRate} and \e{eq:D2}  (dashed-dotted line of light grey (orange online) color ) and given by Eqs.  \e{eq:GrowtRate} and \e{eq:Dlin} (dashed-dotted line of dark grey (brown online) color).  Other lines in Fig. \ref{fig:G_max_vs_phi} use the leading order approximation in $\phi_{eq}$   given  by Eq. \e{gammamax} with four estimates for $\Delta\omega$: from simulations $\Delta\omega=\Delta\omega^{NUM}$; from Eq. \e{eq:dW_Rose} $\Delta\omega=\Delta\omega_{NL}^{Rose}$ and $\Delta\omega=\Delta \omega_{NL}^{Dewar}$  for two cases of Eq. \e{eq:dW_Dewar}.

\begin{figure}
\includegraphics[width=2.8in]{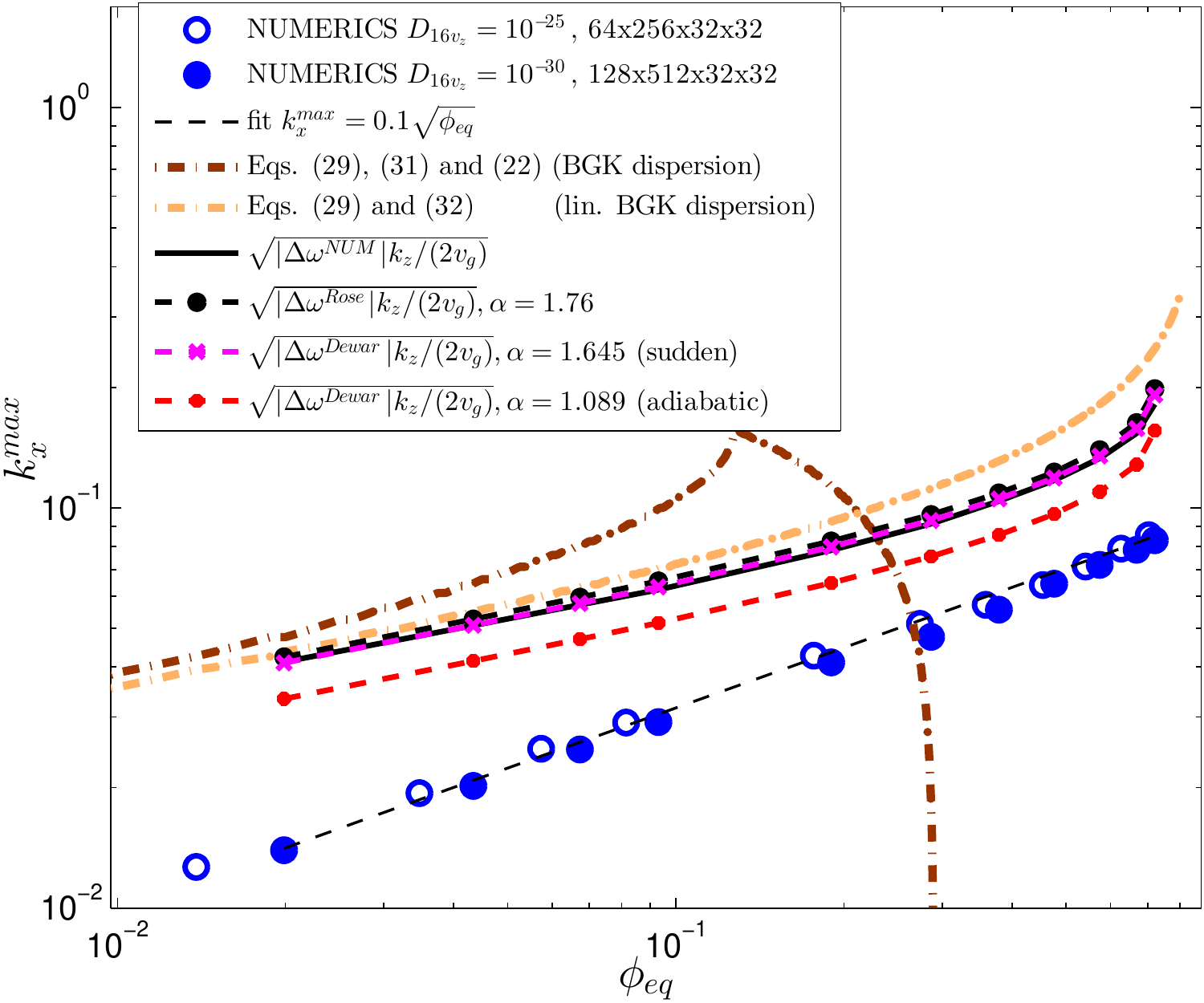}
\caption{(Color online) The wavenumber $k_x^{max}$ at which the growth rate reaches the maximum as a function of BGK amplitude $\phi_{eq}$.}
\label{fig:kx_max_vs_phi}
\end{figure}

We conclude from Fig. \ref{fig:G_max_vs_phi} that while theoretical prediction based on Eqs. \e{eq:Dispersion}, \e{eq:OMEGA}, \e{eq:GrowtRate} and \e{eq:D2}   claims no growth for the amplitudes $\phi_{eq} \gtrsim 0.3,$ we still observe growth for even higher amplitudes. Eqs. \e{eq:GrowtRate} and \e{eq:Dlin} predict growth for any amplitudes but differ from the numerical results by $\sim70\%$ while approximations $\gamma^{max}\approx|\Delta\omega^{Rose}|/4$,$\gamma^{max}\approx|\Delta\omega^{NUM}|/4$ and $\gamma^{max}\approx|\Delta\omega^{Dewar}|/4$ with $\alpha=1.645$ (sudden) work better, especially for amplitudes $\phi_{eq}>0.1$, staying almost identical to each other. While including  $\gamma^{max}\approx|\Delta\omega^{Dewar}|/4$ with $\alpha=1.089$ (adiabatic) curve into Fig. \ref{fig:G_max_vs_phi} for comparison, we believe that it's most appropriate to compare the numerical results to $\gamma^{max}\approx|\Delta\omega^{Dewar}|/4$ with $\alpha=1.645$ (sudden) as Fig. \ref{fig:delta_W_vs_phi} clearly shows that an actual frequency shift $\Delta\omega^{NUM}$ is much closer to $\Delta\omega^{Dewar}$ with ``sudden" distribution rather than ``adiabatic" one. In all these comparisons with theory we assumed in Eq. \e{eq:GrowtRate} that $\nu_{residual}=0$ consistent with the expected absence of sideloss in the periodic BC in $x$ as discussed in the beginning of Section \ref{sec:NUMERICALSIMULATIONS}.
Landau damping, for modes that propagate at some finite angle, is neglected. The authors are not aware of any satisfactory model for such in the literature. That which is available \cite{BergerBrunnerBanksCohenWinjumPOP2015} is ad hoc and fails to properly describe the nonlinear frequency shift. It predicts approximately twice larger nonlinear frequency shift for a wave of given amplitude than the nonlinear frequency shift that Dewar's sudden model or the actual frequency shift of our BGK modes.

The wavenumber $k_x=k_x^{max}$ at which the growth rate has the maximum is shown in Fig. \ref{fig:kx_max_vs_phi} as a function of $\phi_{eq}$ together with the theoretical predictions. Dashed-dotted line of sand color represents prediction by Eqs. \e{eq:Dispersion}, \e{eq:OMEGA}, \e{eq:GrowtRate} and \e{eq:D2}, dashed-dotted line of brown color represents prediction by Eqs.  \e{eq:GrowtRate} and \e{eq:Dlin}, other lines in Fig. \ref{fig:kx_max_vs_phi} use the leading order approximation in $\phi_{eq}$ given  by Eq. \e{kmaxapprox} with the BGK mode group velocity $v_g$ defined in Eq. \e{eq:Dlin}. They include  different estimates of $\Delta\omega$, from Rose's model \e{eq:dW_Rose}, Dewar's model \e{eq:dW_Dewar} and measured $\Delta\omega^{NUM}$. The equation \e{kmaxapprox} predicts $k_x^{max} \propto \sqrt{\Delta\omega}$, which in case of $\Delta\omega\propto\sqrt{\phi_{eq}}$ as in Eqs. \e{eq:dW_Rose} and \e{eq:dW_Dewar} becomes $k_x^{max} \propto (\phi_{eq})^{1/4}$ and fails to agree with numerical results for $k_x^{max}$ somewhat well as seen in Fig. \ref{fig:kx_max_vs_phi}. It is also seen in Fig. \ref{fig:kx_max_vs_phi} that the empirical dependence $k_x^{max} \sim 0.1 \sqrt{\phi_{eq}} $ fits the numerical results pretty well but remains to be explained theoretically.

We also investigated the convergence of growth rates with $D_{16v_z}\rightarrow0$ while $\Delta z,\Delta v_z\rightarrow0$ and, correspondingly, $N_z,N_{v_z}\rightarrow\infty$ while keeping $N_x=32,N_{v_x}=32$ (the discretization in $x$ space does not affect the error in growth rates and 32 points in $v_x$ space together with $v_x^{max}=6$ are enough to resolve the  Maxwellian distribution in $v_x$ direction with error $<10^{-8}$). We found that the relative errors in our numerical results for growth rates with $D_{16v_z}=10^{-30}$ and $128\times 512\times 32\times 32$  grid points for $(z,v_z,x,v_x)$  are within $10-15\%$ range. Reducing $\Delta t$ affected the growth rates results even less so we concluded that $\Delta t=0.1$ was sufficient.

\subsection{Comparison of filamentation instability growth rates with PIC code simulations}
\label{sec:PICComparison}
We now compare   $\gamma_{k_x}$    that we obtained in Section \ref{sec:2DHarveyBGK}  from our simulations for the mode with $(k_z=0.35, k_x=0.05)$ to the growth rates of the same mode obtained using PIC simulations in Fig. 9(j) of Ref. \cite{YinAlbrightBowersDaughtonRosePOP2008}   for three different amplitudes of BGK modes:  $\phi_{eq}=0.2,0.3,0.5$. These BGK modes in both cases were constructed using $k_z=0.35$ and have $v_\varphi=3.35818, \ 3.32288, \ 3.23266$, respectively. Our  growth rates for these three amplitudes are 0.0073,\ 0.0113 and 0.0158.\ The corresponding growth rates from Ref. \cite{YinAlbrightBowersDaughtonRosePOP2008}  are 0.0075, 0.012, and 0.0147, i.e.  only $\sim10\%$ difference with our results.
The total number of particles used in Ref. \cite{YinAlbrightBowersDaughtonRosePOP2008} was $\approx 2\times10^{8}$ with 32x1280 cells and 5000 particles per cell.
Number of grid points in our simulations was 64x256x32x32 for $(z,v_z,x,v_x)$  (total $\approx 1.6\times10^{7}$) with $L_z=2\pi/k_z, L_x=200\pi, v_z^{max}=8,v_x^{max}=6$, and $D_{16v_z}=10^{-25}, \Delta t=0.1, T_{final}=5000$.

$
\\
\\
\\
$

\section{CONCLUSION AND DISCUSSION}
\label{sec:Conclusion}
We studied the linear Langmuir wave (LW) filamentation instability of a particular family of BGK modes that bifurcates from a linear periodic Langmuir wave for $k\lambda_D=0.35$. These BGK modes approximate the nonlinear electron plasma wave resulting from adiabatically slow pumping by SRS. The construction process of these BGK modes is described in detail. Performing direct $2+2D$ Vlasov-Poisson simulations of collisionless plasma we found that the maximal growth rates from simulations are $30-70\%$ smaller compared to the theoretical prediction but exhibit the proper scaling for small amplitudes of BGK wave $\gamma^{max} \propto \sqrt{\phi_{eq}}$ while $k_x^{max} \propto \sqrt{\phi_{eq}}$. These results await an improved theory since current theory predicts $k_x^{max} \propto (\phi_{eq})^{1/4}$.

This behavior contrasts strongly with LW propagation \cite{NicholsonBook1983} in the ``fluid" regime, $k\lambda_D\lesssim 0.2$, in which both two-dimensional (2D) and three-dimensional (3D) collapse \cite{ZakharovJETP1972,ZakharovJETP1972rus,KlineMontgomeryYin2006} may occur if we take into account ion dynamics. Consider a LW wavepacket with electric field amplitude $E$. Its ponderomotive force causes a localized plasma density hole, $\delta n\propto-|E|^2$, which localizes and enhances $|E|$, creating a deeper and narrower hole in the plasma density, and so on, leading to yet larger values of $|E|$ until Landau damping terminates this ``collapse" process.

As shown in Ref. \cite{RosePOP2005}, the transition from the fluid to the regime where the trapped electron effects cannot be ignored occurs at $k\lambda_D\sim 0.2$. Thus at $k\lambda_D\gtrsim0.2$. LW frequency
reduction due to electron trapping may dominate \cite{RosePOP2005} the ponderomotive \cite{NicholsonBook1983} frequency shift \cite{WinjumFahlenMoriPOP2007,ChangDodinPOP2015} with $\triangle  \omega\propto|E|^2$. Contrary to the result of Ref. \cite{WinjumFahlenMoriPOP2007} where fluid nonlinearity frequency shift $\Delta \omega_{fluid}$ is shown to be  positive via use of water bag distribution of electrons, the result of Ref. \cite{ChangDodinPOP2015} indicates that $\Delta \omega_{fluid}$ can have either sign depending on $k$, for example in case of Maxwellian distribution.  Refs. \cite{WinjumFahlenMoriPOP2007} and \cite{ChangDodinPOP2015} suggest that kinetic effects might dominate fluid effects even for large amplitudes of LW if $k\lambda_D>0.3$. Though the trapped electron frequency shift, perturbatively, varies as $|E|^{1/2}$ \cite{ManheimerFlynnPhysFL1971,DewarPhysFL1972,MoralesNeilPRL1972}, and therefore cannot lead to LW collapse \cite{ZakharovJETP1972,ZakharovJETP1972rus,KlineMontgomeryYin2006}, $3D$  PIC simulation results \cite{BowersAlbrightYinBergenKwanPOP2008} have been interpreted as
showing that the trapped electron LW filamentation instability can saturate \cite{YinAlbrightRoseBowersPOP2009}
stimulated Raman back-scatter (SRS)  \cite{GoldmanBoisPhisFL1965} by reducing the LW's coherence.

Since experimental data in the kinetic LW regime is at best qualitative and indirect, such as furnished by observations of SRS light, first principle Vlasov simulations and theory appear to be the chief tools for analyzing LW properties in the kinetic regime. Because LW filamentation is a multi-dimensional effect, with qualitatively different \cite{YinAlbrightRoseBowersPOP2009} 2D versus 3D nonlinear behavior, analysis via Vlasov numerical solutions will remain an outstanding challenge.

\begin{acknowledgments}
This work was supported by the National Science Foundation
under Grants No. PHY 1004118,  No. PHY 1004110 and DMS-1412140.
Simulations were performed at the Center for Advanced Research Computing (CARC) at the University of New Mexico and the Texas Advanced Computing Center (TACC) which was supported by National Science Foundation Grant ACI-1053575.
\end{acknowledgments}



%

\end{document}